\documentclass[prx,aps,showpacs,twocolumn,preprintnumbers,
amsmath,amssymb,superscriptaddress,longbibliography]{revtex4-1}
\usepackage[english]{babel}
\usepackage{amsmath,amssymb,amsfonts}
\usepackage{graphicx}
\usepackage[colorlinks=True,linkcolor=red,citecolor=blue,urlcolor=blue]{hyperref}
\usepackage{bookmark}
\usepackage[dvipsnames]{xcolor} 
\usepackage{braket}
\usepackage{nicefrac}
\usepackage{bm}
\usepackage{bbm}
\usepackage{braket}
\usepackage{slashed}
\usepackage[normalem]{ulem}
\usepackage{array}
\usepackage{makecell}
\newcolumntype{C}{>{$}c<{$}}

\renewcommand{\Re}{\mathop{\mathrm{Re}}}
\renewcommand{\Im}{\mathop{\mathrm{Im}}}

\renewcommand{\i}{\mathrm{i}}

\allowdisplaybreaks
\DeclareMathAlphabet{\zc}{OT1}{pzc}{m}{it}

  \DeclareUnicodeCharacter{2212}{-}

\begin{document}

\title{Topological phase diagrams of exactly solvable \\
non-Hermitian interacting Kitaev chains}
\author{Sharareh Sayyad}
\email{sharareh.sayyad@mpl.mpg.de}
\affiliation{Max Planck Institute for the Science of Light, Staudtstra\ss e 2, 91058 Erlangen, Germany}
\author{Jose L. Lado}
\affiliation{Department of Applied Physics, Aalto University, FI-00076 Aalto, Espoo, Finland}

\date{\today}

\begin{abstract}
    Many-body interactions give rise to the appearance of exotic phases in Hermitian physics. Despite their importance, many-body effects remain an open problem in non-Hermitian physics due to the complexity of treating many-body interactions. Here, we present a family of exact and numerical phase diagrams for non-Hermitian interacting Kitaev chains. In particular, we establish the exact phase boundaries for the dimerized Kitaev-Hubbard chain with complex-valued Hubbard interactions. Our results reveal that some of the Hermitian phases disappear as non-Hermiticty is enhanced. 
    Based on our analytical findings, we explore the regime of the model that goes beyond the solvable regime, revealing regimes where non-Hermitian topological degeneracy remains. The combination of our exact and numerical phase diagrams provides an extensive description of a family of non-Hermitian interacting models. Our results provide a stepping stone toward characterizing non-Hermitian topology in realistic interacting quantum many-body systems.
\end{abstract}

\maketitle

\paragraph*{\bf Introduction.}
Many-body interactions play a crucial role in Hermitian quantum systems. The emergent correlation effects in these systems give rise to a variety of collective phenomena, such as spontaneous symmetry breaking~\cite{Koma1994, Belyaev2006, Brauner2010, Dong2017}, phase transitions~\cite{Santos2016, Heyl2018, Carollo2020, Serwatka2023}, and the emergence of fractionalized quasiparticles~\cite{Zhang2020, Hashisaka2021, Kaskela2021, Wouters2022}. Understanding these rich phenomena often requires
the combination of analytical and numerical techniques
due to the scarcity of exact solutions for many-body models.
Nonetheless, especially in one-dimensional systems, analytical solutions in specific regimes are attainable~\cite{Gangadharaiah2011, Katsura2015, Ezawa2017, Wang2017, Miao2017, Zvyagin2022}. Away from these parameter regimes, employing various numerical methods~\cite{Lin1990, Vidal2004, Schollwock2005, Kotliar2006, Gangadharaiah2011, Stoudenmire2011, Silvi2019, Tuovinen2021} allows unveiling the underlying physics of many-body systems in generic scenarios.

The presence of losses and dissipation in real systems provide natural platforms
realizing non-Hermitian models~\cite{Esaki2011, Malzard2015, Gong2018, Yao2018, Shen2018, Kawabata2019, Yokomizo2019, Zhou2019, Borgnia2020, Ashida2020}. Non-Hermitian quantum models have risen as a new paradigm to manipulate and interpret various emergent phenomena~\cite{Sayyad2022d, Okuma2023}. Here, Non-Hermiticity emerges as the effective description~\cite{Prosen2008, Lieu2020, Sayyad2021,  Perina2022, Yang2022, Talkington2022, Starchl2022, Gneiting2022} of out-of-equilibrium and open quantum systems, e.g., in superconducting qubits~\cite{Chen2021, Abbasi2022, Chen2022}, giving rise to various phenomena absent in the Hermitian counterpart. Paradigmatic examples are occurrence of various non-Hermitian degeneracies~\cite{Leykam2017, Bergholtz2021, Sayyad2022, Sayyad2022b, Sayyad2022c} and non-Hermitian (bulk) skin states~\cite{Song2019, Zhang2022}. Both of these phenomenologies are mainly explored in effectively single-particle non-Hermitian Hamiltonians while non-Hermitian many-body effects have remained relatively unexplored, partly due to limitations of numerical methods\cite{Freund1992, Freund1993, Guo2022, Carden2011, Zhang2016z, Chen2022a}.
In particular, recent efforts have addressed one-dimensional non-Hermitian fermionic~\cite{Fukui1998, Buca2020, Zhang2021, Nakagawa2021, Yoshida2022, Hyart2022} and bosonic~\cite{Yamamoto2022, Wang2022} Hubbard models. Here, the non-Hermiticity is incorporated by having nonreciprocal hopping~\cite{Fukui1998}, complex Hopping~\cite{Zhang2021}, or complex Hubbard interaction~\cite{Buca2020,Nakagawa2021,Yoshida2022}. The latter form of non-Hermiticity provides effective descriptions for experiments on open quantum systems with two-body loss~\cite{Lewenstein2007, Zhang2016, Gross2017, Rosso2022}.

In this Letter, combining exact analytical results and numerical calculations, we establish
the phase diagram of a family of non-Hermitian Kitaev chains~\cite{Rainis2012, Lieu2019, Sayyad2021, Sakaguchi2022}. Our interacting model consists of a complex-valued many-body interaction that may host Majorana modes, in particular, realizable in an array of Josephson junctions~\cite{Hassler2012}. Our results reveal that, depending on the relative couplings, non-Hermitian interacting phases with topological degeneracies emerge in the system. We show how increasing the non-Hermiticity parameters affect some of the many-body topological phases. We furthermore present that the topological degeneracies of the model remain in the non-analytically solvable regime by numerically solving the interacting problem.

\begin{figure}[t!]
    \centering
    \includegraphics[width=0.9\columnwidth]{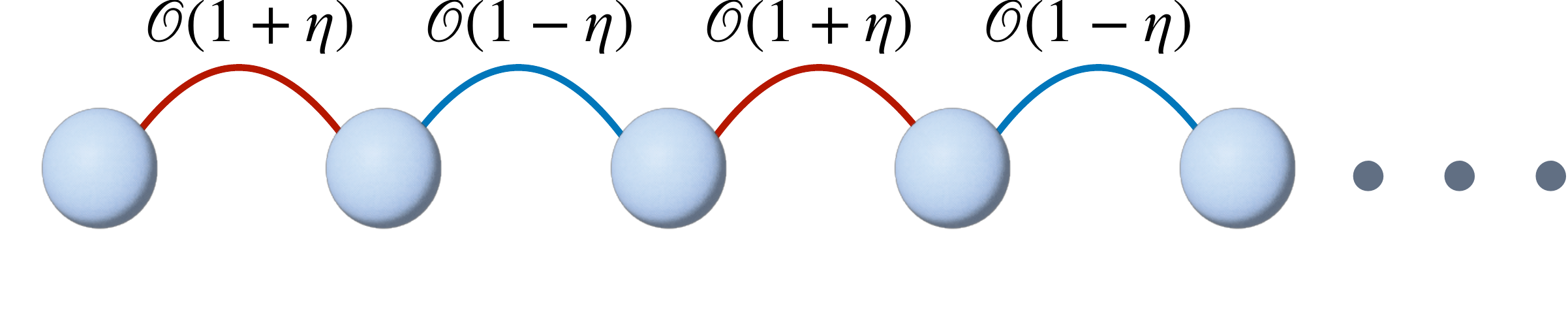}
    \caption{
    Schematic illustration of a 1D dimerized Kitaev-Hubbard chain with dimerized hopping~$t$, pairing~$\Delta$, and complex-valued Hubbard interaction~$U-\i \delta$. Here, $\eta$ denotes the dimerization parameter and ${\cal O} \in \{ t, \Delta, U, \delta \}$.
    \label{fig:sketch}
    }
\end{figure}

\begin{figure}[ht!]
    \centering
    \includegraphics[width=\columnwidth]{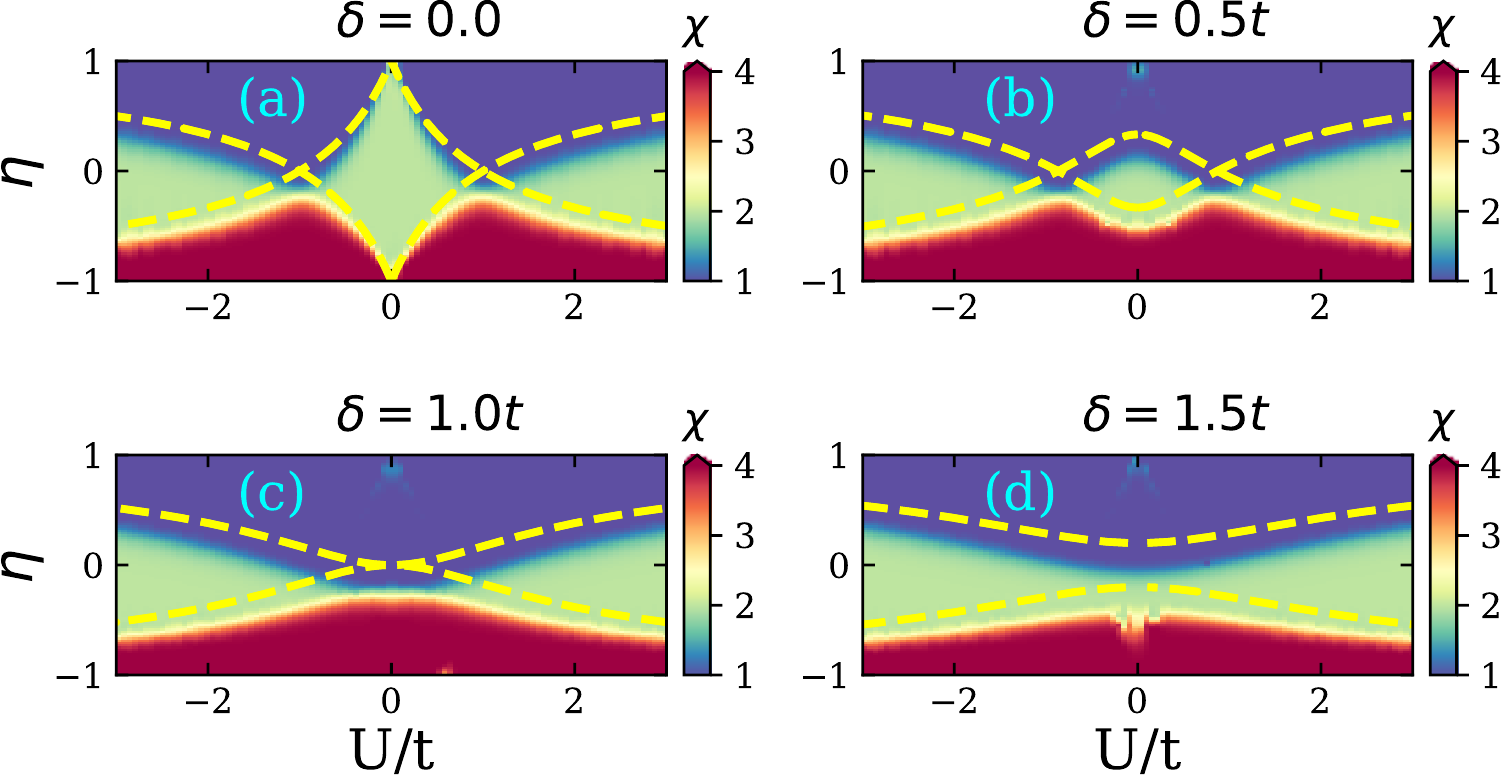}
    \caption{
    Phase diagrams of the non-Hermitian Hamiltonian and associated orders of degeneracies on the $(U/t- \eta)$ plane at $\mu=0$ and $\Delta=t$. Non-Hermiticity parameter is set to $\delta/t=$ $0$~(a), $0.5$~(b), $1.0$~(c), $1.5$~(d). 
    The yellow dashed lines display the exact phase boundaries given by Eq.~\eqref{eq:phasecond0}. The heat map shows the degeneracies obtained numerically for a chain with length $L=16$, shown in red, green, and blue, respectively, for fourfold, twofold, and no degeneracies.  
    \label{fig:phasediag}
    }
\end{figure}

\paragraph*{\bf Model.}
The Hamiltonian for the non-Hermitian dimerized Kitaev-Hubbard chain, schematically shown in Fig.~\ref{fig:sketch}, reads
\begin{align}
  & {\cal H} =
        - \sum_{j=1}^{L-1}
        \left[ 
        t_{j} \left(   c^{\dagger}_{j} c_{j+1} + 
     c^{\dagger}_{j+1} c_{j} \right)
     +  \Delta_{j}  \left( c^{\dagger}_{j}c^{\dagger}_{j+1} +  c_{j+1}c_{j}  \right)
     \right]
         \nonumber \\
    &
    + \sum_{j=1}^{L-1} (U_{j} - \i \delta_{j} ) \left( 2n_{j} -1 \right) \left(2 n_{j+1} -1 \right)
    - \mu \sum_{j=1}^{L}  (n_{j}-\frac{1}{2} )
     ,
     \label{eq:Hintsshk}
\end{align}
where $L$ denotes the length of the chain, $c^{\dagger}_{j}~(c_{j})$ creates~(annihilates) an spinless fermion at site $j$ associated with the fermion density $n_{j}=c^{\dagger}_{j} c_{j}$.
Here $\mu$ adjusts the onsite energy, and $t_{j}$, $\Delta_{j}$ and $U_{j}$ are, respectively, real-valued site-dependent hopping amplitude, superconducting pairing amplitude, and Hubbard interaction. 
The dimerized parameter ${\cal O}_{j} \in \{ t_{j}, \Delta_{j},  U_{j}, \delta_{j} \}$ for $1 \leq j \leq L$ reads
\begin{align}
    {\cal O}_{j} &= 
    \begin{cases}
    {\cal O} (1- \eta), \quad \, j \,{\rm mod}\,2 = 0,\\
    {\cal O} (1+ \eta), \quad  \, j \, {\rm mod}\,2 = 1,
    \end{cases}
\end{align}
where $\eta$ is the real-valued dimerization parameter and ${\cal O} \in \{t, \Delta, U, \delta \}$ stands for site-independent parameters; see also Fig.~\ref{fig:sketch}.

The Hamiltonian in Eq.~\eqref{eq:Hintsshk} at $\mu=0$ remains invariant under $c_{j} \rightarrow (-1)^{j} c^{\dagger}_{j}$ which enforces the charge conjugation symmetry. We note that respecting this symmetry ensures that eigenvalues of the Hamiltonian come in complex-conjugate pairs making this model not directly realizable in open quantum systems~\cite{Lieu2020}, which can be resolved by a negative imaginary shift of all eigenvalues~\cite{Joglekar2018}. Nevertheless, preserving the charge conjugation symmetry in Hermitian models~($\delta=0$) at the symmetric point $\Delta=t$ allows calculating exact phase diagrams for arbitrary $\eta$~\cite{Gangadharaiah2011, Katsura2015, Ezawa2017, Wang2017, Miao2017}.
In the following, we present that there exists an analytical solution for the non-Hermitian model when the charge conjugation symmetry is respected, i.e., at $\mu=0$ and $\Delta=t$. When $\Delta \neq t$ or $\mu \neq 0$, we compute the topological phase diagram numerically.

\begin{figure}[t]
    \centering
    \includegraphics[width=\columnwidth]{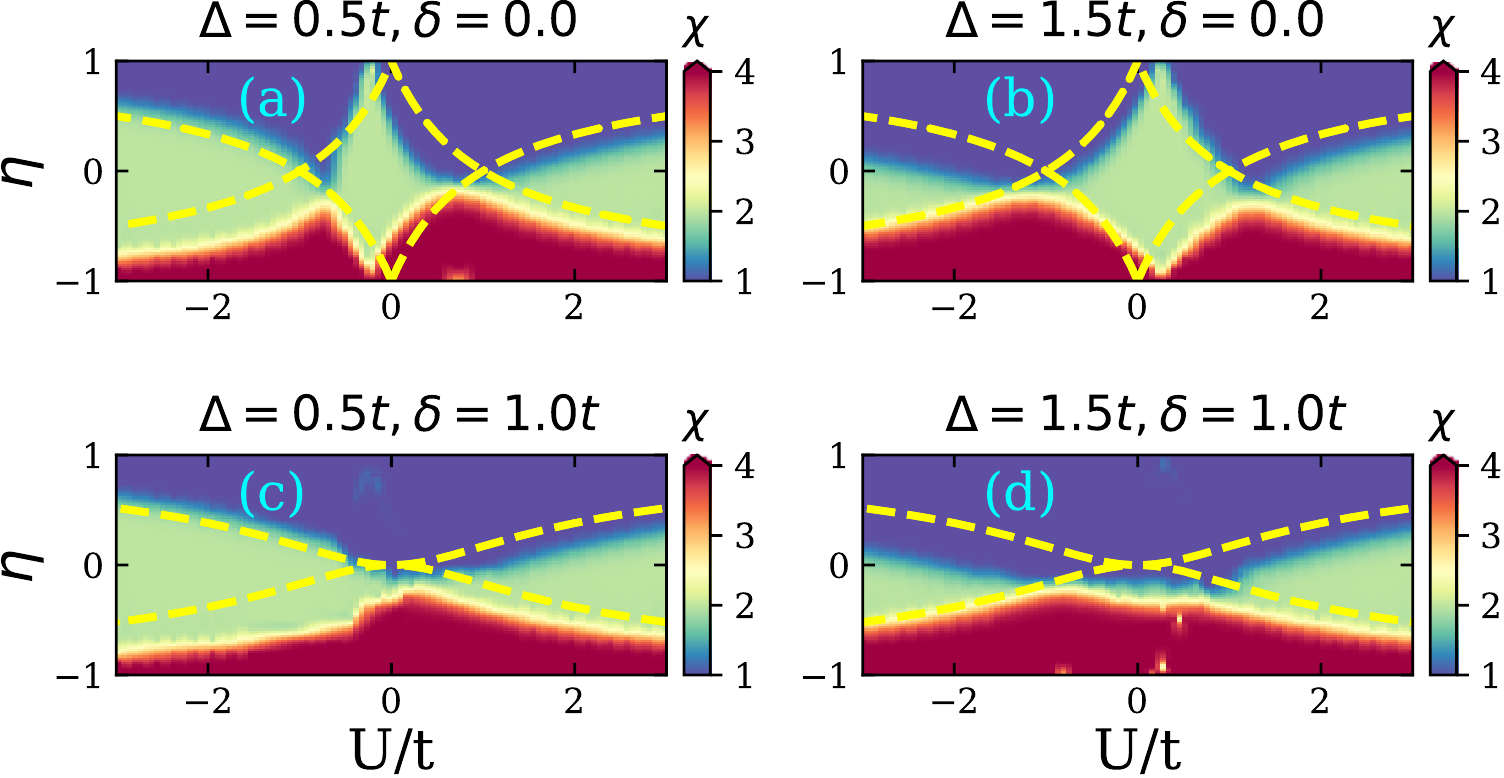}
    \caption{
    Phase diagrams of the non-Hermitian Hamiltonian and associated orders of degeneracies on the $(U/t- \eta)$ plane at $\mu=0$ and $\Delta\neq t$.
    The heat map corresponds to the degeneracies obtained numerically for a system with $L=16$. The superconducting pairing amplitude and non-Hermiticity are set to $(\Delta/t, \delta)=$ $(0.5,0.0)$~(a), $(1.5,0.0)$~(b), $(0.5,1.0)$~(c), $(1.5,1.0)$~(d). The yellow dashed lines are phase boundaries at $\Delta=t$ and are shown as a guide for eye.
    \label{fig:phasediag_dpair}
    }
\end{figure}

\paragraph*{\bf Exact solution at $\mu=0$ and $\Delta=t$.}

 To obtain the exact phase diagram of the model Hamiltonian at $\mu=0$, we employ two Jordan-Wigner transformations and one spin rotation; see the Supplemental materials~(SM) for details~\cite{SuppMat}. This procedure map our initial non-Hermitian interacting Hamiltonian into a non-Hermitian quadratic fermionic model given by
 \begin{align}
    {\cal H} &= \sum_{j} - t_{j} \left[  f^{\dagger}_{j+1}f_{j} +f^{\dagger}_{j}  f_{j+1}
    + f^{\dagger}_{j} f^{\dagger}_{j+1}  + f_{j+1} f_{j} \right]\nonumber \\
    &+\sum_{j} \tilde{U}_{j} \left[ f^{\dagger}_{j+1}f_{j} +f^{\dagger}_{j}  f_{j+1}
    - f^{\dagger}_{j} f^{\dagger}_{j+1}  - f_{j+1} f_{j} \right],
    \label{eq:Hscf1}
\end{align}
which in the momentum-space casts
 \begin{align}
    {\cal H} &=  \sum_{k} (-t +\tilde{U}) \left[  z_{k} f^{\dagger}_{k B} f_{kA} + z_{k}^{*} f^{\dagger}_{k A} f_{k B} \right]
    \nonumber \\
    &-\sum_{k} (t +\tilde{U}) \left[ 
    w_{k} f^{\dagger}_{-k A} f^{\dagger}_{kB} 
    + w_{k}^{*} f_{k B} f_{-k A} 
    \right]
    ,
    \label{eq:Hscfk0}
\end{align}
where $\tilde{U}=U-\i \delta$.
Diagonalizing this Hamiltonian, we obtain the energy spectrum of this four-band system given by
\begin{align}
    \frac{\Lambda_{k}^2}{4} &=
    \begin{cases}
    \tilde{U}^{2} (1 + \eta)^{2} + t^2 (1-\eta)^{2} -2 t \tilde{U}(1-\eta^{2}) \cos(k)
    ,\\
    \tilde{U}^{2} (1 - \eta)^{2} + t^2 (1+\eta)^{2} -2 t \tilde{U}(1-\eta^{2}) \cos(k)
,
    \end{cases}
    \label{eq:Lameig0}
\end{align}
where $\tilde{U}^{2} = U^{2} -\delta^{2} - 2 \i \delta U$. As eigenvalues of our system appear in complex-conjugate pairs, due to the charge-conjugation, zero modes in $\Lambda_{k}$ emerge when $\Re[\Lambda_{k}]=0$ at
\begin{align}
    \frac{U}{t}=    \sqrt{\frac{\delta^2}{t^2} - \frac{(1 \pm \eta)^2}{(1 \mp \eta)^2 }},
    \label{eq:phasecond0}
\end{align}
which is obtained at $k=\pm \pi/2$. In the Hermitian limit~($\delta=0$), Eq.~\eqref{eq:phasecond0} reproduces the Hermitian results~\cite{Gangadharaiah2011, Katsura2015, Ezawa2017,Wang2017,Miao2017}. Eq.~\eqref{eq:phasecond0} shows the boundaries between various phases in our system, presented in yellow dashed lines in Fig.~\ref{fig:phasediag_dpair}. These boundaries delineate phases with fourfold, twofold, and no degeneracies, respectively, shown in red, green, and blue in Fig.~\ref{fig:phasediag}. 
The order of degeneracy is determined by 
$\chi= \sum_{i=0}^{L} \exp[- \lambda |\varepsilon_{i}-\varepsilon_{0}|]$. This quantity measures the degeneracy between first~$\varepsilon_{0}$ and the $i$the smallest eigenvalue~$\varepsilon_{i}$ within the energy resolution of $1/\lambda (=0.05)$. Here the eigenvalues are calculated using the numerical exact diagonalization method. The difference between the numerical boundaries and the exact solution should be attributed to the finite size effect. In the thermodynamic limit~($L=\infty$), one can recover the exact phase boundaries; see the SM~\cite{SuppMat}.
The topological superconducting phase residing in the boundaries surrounding $U=0$~\cite{Kitaev2001} shrinks as non-Hermiticty increases. At the critical value $\delta=t$, this topological phase fades away, resulting in the mixing of other two-fold degenerate phases. 
It is also worth noting that the exact phase boundaries in Eq.~\eqref{eq:phasecond0} are associated with transitions between non-Hermitian spectra with different types of the gap in the non-Hermitian effective model in Eq.~\eqref{eq:Hscfk0}; see also the SM~\cite{SuppMat}.

\begin{figure}[t]
    \centering
    \includegraphics[width=\columnwidth]{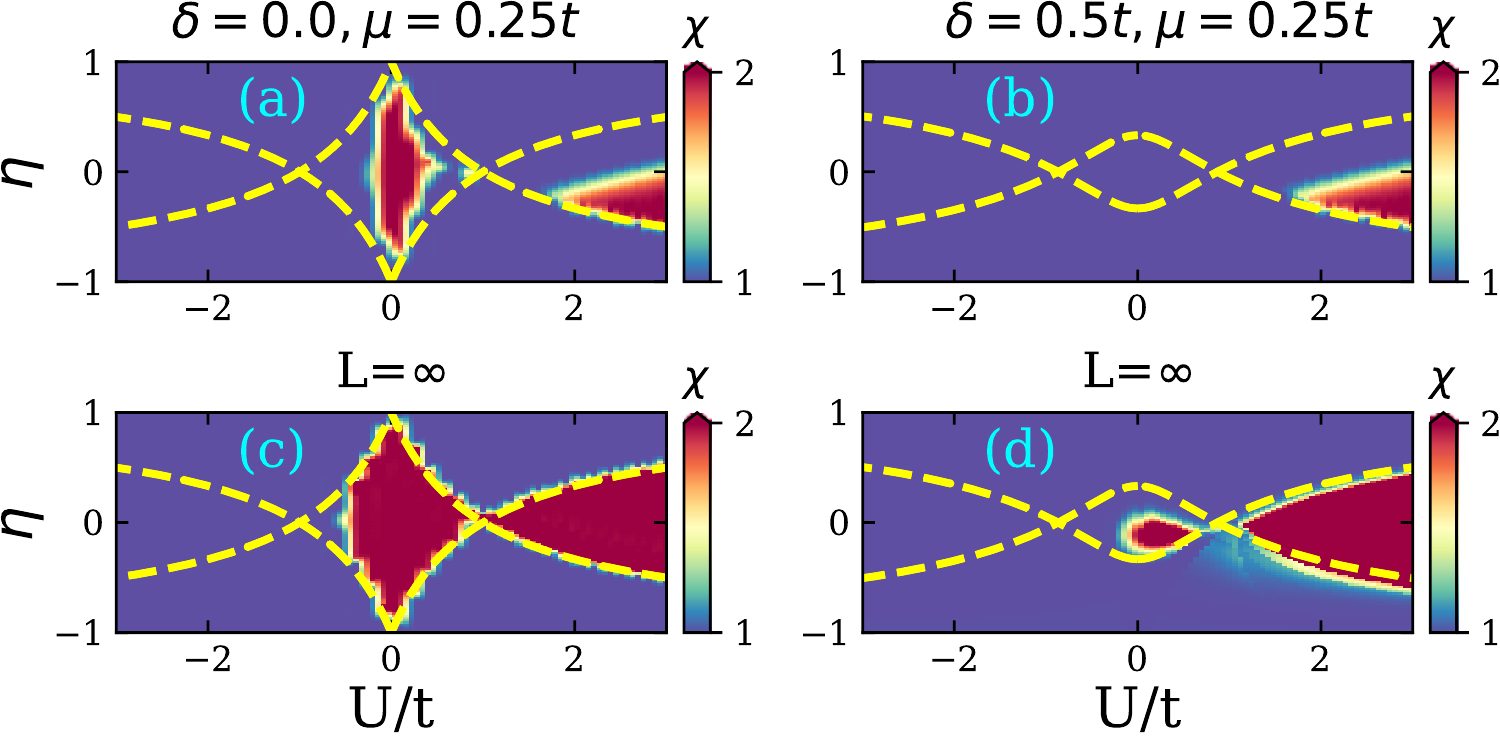}
    \caption{
    Phase diagrams of the non-Hermitian Hamiltonian and associated orders of degeneracies on the $(U/t- \eta)$ plane with $\mu \neq 0$. The superconducting pairing amplitude is set to $\Delta = t$. 
    The chain size is set to $L=16$ in (a,b), and extrapolated data in the thermodynamic limit are shown in (c,d). The yellow dashed lines are phase boundaries at $\mu=0$ and are shown as a guide for the eye.
    We took $\delta,\mu=0,0.25t$ in (a,c) and $\delta,\mu=0.5,0.25t$ in (b,d)
    \label{fig:phasediag_mu}
    }
\end{figure}

\paragraph*{\bf Many-body Majorana edge modes.}
To identify Majorana modes in our model, in the next step, we rewrite the Hamiltonian in Eq.~\eqref{eq:Hscf1} in terms of Majorana fermions ${\Upsilon}_{j}^{A} = f^{\dagger}_{j} + f_{j}$ and ${\Upsilon}_{j}^{B} = \i (f^{\dagger}_{j} -f_{j})$.The Hamiltonian then reads
\begin{align}
    {\cal H} = \i \sum_{j}  \left[ 
             t_{j} {\Upsilon}_{j}^{B} \Upsilon_{j+1}^{A}
           + \tilde{U}_{j} \Upsilon_{j}^{A} \Upsilon_{j+1}^{B}
             \right].
             \label{eq:Hmaj}
\end{align}
We note that interactions and hopping amplitudes between the same sublattices or within each unit cell vanish. Hence, Eq.~\eqref{eq:Hmaj} can be decoupled into two independent non-interacting Kitaev chains with length $L/2$ such that ${\cal H}={\cal H}_{\rm I}+{\cal H}_{\rm II}$ where
\begin{align}
    {\cal H}_{\rm I} 
    &=
    \sum_{j=1}^{L/2} 
    \left[ -\i t_{2j} 
    \Phi_{{\rm I},j+1}^{A} \Phi_{{\rm I},j}^{B}
    +\i \tilde{U}_{2j-1} 
    \Phi_{{\rm I},j}^{A} \Phi_{{\rm I},j}^{B}
    \right]
    \label{eq:HeffI}
    ,\\
    {\cal H}_{\rm II} 
    &=
    \sum_{j=1}^{L/2}
    \left[ 
    \i \tilde{U}_{2j} 
    \Phi_{{\rm II},j+1}^{B} \Phi_{{\rm II},j}^{A}
    -\i t_{2j-1} 
    \Phi_{{\rm II},j}^{B} \Phi_{{\rm II},j}^{A}
    \right]
     \label{eq:HeffII}
    ,
\end{align}
with $\Phi^{A}_{{\rm I},j}= \Upsilon_{2j-1}^{A}$, $\Phi^{B}_{{\rm I},j}=\Upsilon^{B}_{2j}$, $\Phi^{B}_{{\rm II},j}=\Upsilon^{B}_{2j-1}$, and $\Phi^{B}_{{\rm II},j}=\Upsilon^{A}_{2j}$.
Introducing Majorana particles from the electron operators as $\gamma_{j}^{A} = c^{\dagger}_{j} + c_{j}$ and $\gamma_{j}^{B} = \i (c^{\dagger}_{j} -c_{j})$, one can show that ${\Upsilon}$ operators are products of $\gamma$ operators such that~\cite{Goldstein2012,Yang2014,Kells2015,Miao2017,McGinley2017,Ezawa2017}
\begin{align}
    \Upsilon_{j}^{A} &=
    \begin{cases}
      \prod^{j-1}_{k=\text{odd}} 
    \left[ \i \gamma_{k}^{B} \gamma_{k+1}^{A} \right] \gamma_{j}^{A} , \quad &j=\text{odd},\\
    &\\
     \prod_{k=\text{odd}}^{j-3} \left[ \i \gamma_{k}^{A} \gamma_{k+1}^{B}\right]
    (\i \gamma^{A}_{j-1} \gamma_{j}^{A}), \quad &j=\text{even},\\
    \end{cases}
\end{align}
\begin{align}
        \Upsilon_{j}^{B} &=
    \begin{cases}
      \prod^{j-2}_{k=\text{odd}} 
    \left[ \i \gamma_{k}^{A} \gamma_{k+1}^{B} \right] \i \gamma_{j}^{A} \gamma^{B}_{j} , \quad &j=\text{odd},\\
    &\\
     \prod_{k=\text{odd}}^{j-1} \left[ \i \gamma_{k}^{B} \gamma_{k+1}^{A}\right]
    \gamma^{B}_{j} , \quad &j=\text{even}.
    \end{cases}
\end{align}
These relations keep the Majorana anti-commutation relations unchanged, i.e., $ \{\Upsilon^{\alpha}_{i}, \Upsilon^{\beta}_{j}\}=\{\gamma^{\alpha}_{i},\gamma^{\beta}_{j}\}=2 \delta_{i,j}\delta^{\alpha, \beta}$.
We note that $\Upsilon$ operators, which are comprised of odd~(even) number of Majorana fermions~($\gamma$s), belong to subsystem ${\rm I}~({\rm II})$ described by ${\cal H}_{\rm I}~({\cal H}_{\rm II})$.

\begin{figure*}[t]
    \centering
    \includegraphics[width=0.99\textwidth]{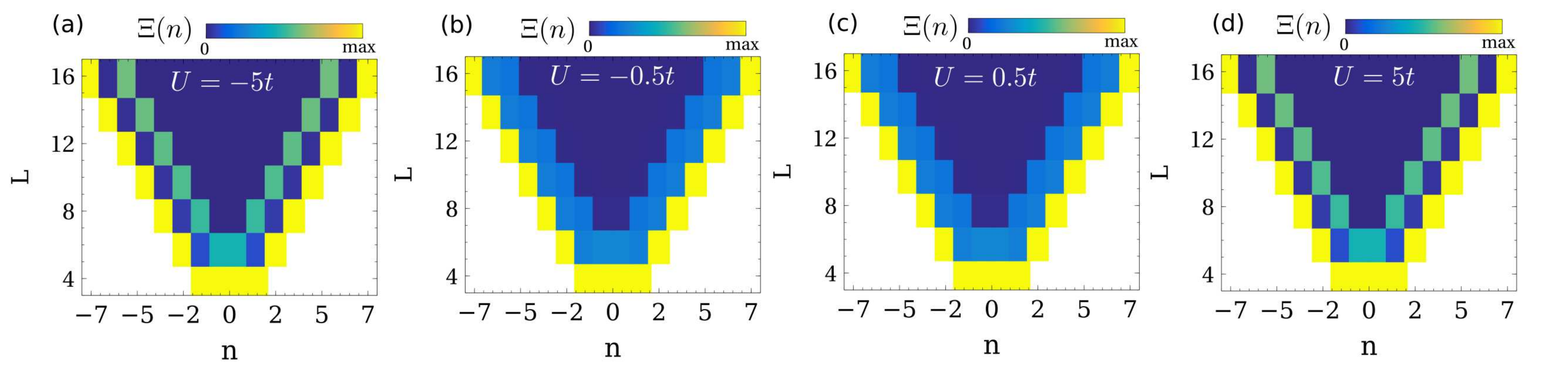}
    \caption{
    Spatial distribution of the zero modes for different system lengths
    for the non-Hermitian model computed with the local response function
    given by Eq.~\eqref{eq:local}. It is observed that both in the presence of repulsive and attractive interactions, an edge response appears, accounting for the topological degeneracy of the model. The edge excitations emerge both in the case of two-fold (a,d) and four-fold (b,c) degeneracy.
    We took $\mu=0$, $\Delta=t$, $\eta=-0.6$ and $\delta=0.5t$. 
    \label{fig:local}
    }
\end{figure*}

The quadratic Hamiltonian in Eq.~\eqref{eq:Hmaj} may host two types of boundary modes~($\cal Q$). These boundary modes are constructed from linear combinations of $\Phi^{ A/B}_{\rm I/II}$ operators, i.e., ${\cal Q}^{\alpha}_{\beta}= \sum_{j \geq 0} { a}_{\beta, j} \Phi^{\alpha}_{\beta,j}$ with $\alpha=A,B$ and $\beta= {\rm I,II}$. As ${\cal Q}$ consists of higher-order multiple Majorana fermions, it is dubbed "many-body Majorana operator"~\cite{McGinley2017}.
In the Hermitian limit, ${\cal Q}$ operators are conserved, $[{\cal H}, {\cal Q}]=0$, and using the iteration procedure, one can determine the coefficients~($a$) as~\cite{Fendley2012, Ezawa2017, McGinley2017}
\begin{align}
    { a}_{{\rm I}, j}= 
    - \left( \frac{{U} (1+ \eta)}{ t(1-\eta)} \right)^{j-1}
    ,
    \quad 
     { a}_{{\rm II}, j}= 
     - \left( \frac{t (1+ \eta)}{ {U}(1-\eta)} \right)^{j-1}
     .
     \label{eq:acoeff}
\end{align}
In non-Hermitian systems, the operator $\cal O$ is conserved if it satisfies $[{\cal H}_{R}, {\cal O}]= \{{\cal H}_{I}, {\cal O}\} = 0$ where ${\cal H}= {\cal H}_{R} + \i {\cal H}_{I}$~\cite{Sayyad2022c}. We note that based on the structure of ${\cal H}_{\rm I/II}$ in Eqs.~\eqref{eq:HeffI} and \eqref{eq:HeffII}, $\{{\cal H}_{I}, {\cal O}\}=0 $ is by construction satisfied and fulfilling $[{\cal H}_{R}, {\cal O}]=0$ results in obtaining Eq.~\eqref{eq:acoeff}. Hence, the boundary modes in our non-Hermitian system are continuously~($\delta \rightarrow 0$) connected to the zero-energy boundary modes.

The Majorana boundary mode ${\cal Q}_{\rm I}$ consists of odd numbers of higher-order Majorana operators~($\gamma$) is fermionic and satisfies $\{ {\cal Z}_{2}, {\cal Q}_{\rm I} \}=0$, in the infinite chain limit~\cite{Fendley2012}. However,  the ${\cal Q}_{\rm II}$ mode comprises even numbers of higher-order Majorana operators~($\gamma$) is bosonic as $[ {\cal Z}_{2}, {\cal Q}_{\rm II} ]=0$~\cite{Katsura2015, McGinley2017, Ezawa2017}. Regions with fourfold degeneracies in Fig.~\ref{fig:phasediag} host both $({\cal Q}_{\rm I} ,{\cal Q}_{\rm II} )$. The topological superconducting phase enclosing $U=0$ merely hosts ${\cal Q}_{\rm I} $, in agreement with Hermitian noninteracting intuition~\cite{Kitaev2001}. The other two twofold degenerate phases accommodate ${\cal Q}_{\rm II}$.

\paragraph*{\bf Beyond exact solutions. }

Let us now present the phase diagram of our system away from the integrable regime $\Delta=t$ and $\mu=0$. First, we consider $\Delta/t \in \{0.5, 1.5\}$ and plot the associated phase diagrams with $\delta \in \{0.0, 1.0\}$ in Fig.~\ref{fig:phasediag_dpair} for a finite size system with $L=16$. Similar to the phase diagram of the system at $\Delta=t$, we witness phases with fourfold, twofold, and no degeneracies. Comparing the exact phase boundaries at $\Delta=t$, in yellow dashed lines, with boundaries of regions with $2n$fold degeneracies, in red or green, we identify deformation of the phase boundaries toward $U<0$~($>0$) for $\Delta <t~(>t)$. 

At finite chemical potential and $\delta=0$, all fourfold degeneracies are lifted, and merely phases with twofold degeneracies remain in the phase diagram; see Fig.~\ref{fig:phasediag_mu}. The Hermitian phase boundary $U=-0.5 t$ at $\eta=0$ is consistent with previous calculations on the Kitaev-Hubbard chain~\cite{Mahyaeh2020}; see panels (a) at $L=16$, (c) in the thermodynamics limit and the SM~\cite{SuppMat}. Witnessing merely twofold degeneracies in phases with $U>1$ in the Hermitian limit also persists as non-Hermiticty is increased; see Fig.~\ref{fig:phasediag_mu}(b) at $L=16$. While no portion of the topological superconducting phase, region encircling $U=0$, is present in (b) obtained at $\mu=0.25$ and $L=16$, extrapolating the phase diagram using different system sizes, shown in (d), reveals the survival of this phase at $L=\infty$.

\paragraph*{\bf Experimental measurement. }

We now address the signatures of the zero modes from the experimental point of view.
In a tunneling experiment with a local probe~\cite{Drost2017,  Kempkes2019, Kempkes2018, Huda2020, Dvir2023}, the conductance
at zero bias $G(\omega=0,n)$ depends on the probability of extracting (injecting) an electron in site $n$
at energy $\omega$ as $\text{dI/dV} (n,\omega) \sim \sum_\alpha |\langle GS |c^{(\dagger)}_n | \Psi_\alpha \rangle|^2\delta(\omega - E_\alpha + E_{GS})$,
where $|GS\rangle $ is the ground state and $H|\Psi_\alpha\rangle = E_\alpha |\Psi_\alpha \rangle$ are the many-body
excited states of the system. In the presence of topological degeneracy, the ground state presents zero mode
excitations that distinguish different ground states. In that scenario, the zero bias conductance at site $n$
can be written as $\text{dI/dV} (n,\omega=0) \sim \Xi(n)$ where

\begin{equation}
    \Xi(n) = \sum_{\alpha}
    |\langle \Psi_\alpha | c_n | GS \rangle|^2 +
    |\langle \Psi_\alpha | c^\dagger_n | GS \rangle|^2
    \label{eq:local}
\end{equation}

which directly images the probability of a local excitation between the ground state
and its degenerate manifold. Here, $\alpha$ runs over
the ground state manifold. In particular, $\Xi(n)$ allows one to directly
observe the emergence of topological zero modes associated with the
topological degeneracy of the ground state.
With the previous quantity, the emergence of topological zero modes associated to the
topological degeneracy of the non-Hermitian model can be directly imaged.
We show in Fig.~\ref{fig:local} the local correlator computed
for the interacting non-Hermitian model for different system sizes.
As the system becomes larger, an edge excitation emerges in the model,
which in the thermodynamic limit leads to decoupled modes between the two edges; see also the SM~\cite{SuppMat}.
The previous quantity has been directly imaged in the realization of the current model in the topological phase with $\delta=U=0$ and $\eta \neq 0$~\cite{Drost2017}, and minimal a chain with $\Delta \neq 0$\cite{Dvir2023}.

\paragraph*{\bf Conclusion. }
To summarize, we have presented a family of non-Hermitian interacting models featuring
different classes of topological degeneracies. 
While non-interacting non-Hermitian models can be studied with conventional methodologies,
the inclusion of many-body interactions renders exploring non-Hermitian systems greatly challenging.
Our manuscript establishes a family of solvable interacting non-Hermitian models, providing ideal
systems for benchmarking methodologies to treat interacting non-Hermitian models.
Besides showing the emergence of different topological phases in the solvable limit, we provided the many-body operators accounting for the topological degeneracy of the model.
We showed how non-Hermiticity modifies topological many-body models, substantially impacting the topological phases of Hermitian systems.   
We benchmarked our analytical construction with exact numerical calculations of the full many-body system, demonstrating that even for finite systems, the emergence of topological modes and topological
degeneracies can be observed.
Our results establish a versatile family of models featuring interacting topology,
providing a starting point for higher dimensional solvable interacting models.

\paragraph*{\bf Acknowledgement. }
S.S. thanks Vittorio Paeno and Abolhassan Vaezi for helpful discussions.
J.L.L. acknowledges
the computational resources provided by
the Aalto Science-IT project,
and the
financial support from the
Academy of Finland Projects No.
331342 and No. 336243
and the Jane and Aatos Erkko Foundation.

\appendix
\renewcommand{\theparagraph}{\bf \thesubsubsection.\arabic{paragraph}}

\renewcommand{\thefigure}{S\arabic{figure}}
\setcounter{figure}{0} 

\renewcommand{\theequation}{S\arabic{equation}}
\setcounter{equation}{0} 
\section{Mapping the non-Hermitian interacting Kitaev chain into a noninteracting Hamiltonian}

Following Refs.~\cite{Ezawa2017, Wang2017}, we now present the generalization of mapping our non-Hermitian interacting model into a non-interacting Hamiltonian when $t=\Delta$ and $\mu=0$.

In the first step, we represent the fermion operators in terms of the spin operators, described by Pauli matrices, using the Jordan-Wigner transformation~\cite{Sela2011, Katsura2015, Miao2017, Mahyaeh2020, Liu2021, Bi2021, Zvyagin2022} given by $c_{i}  =K_{i} \sigma^{-}_{i}$, $ c_{i}^{\dagger} = \sigma^{+}_{i} K^{+}_{i} $, and $2n_{i}-1 = \sigma_{i}^{z}$ with $\sigma^{\pm}=\sigma^{x} \pm \i \sigma^{y}$ and $K_{i} = \prod_{j<i} (-\sigma^{z}_{j})$.
As a result, the spin representation of the Hamiltonian in Eq.~(1) casts
\begin{equation}
    {\cal H} = \sum_{j} \left[ 
    -t_{j} \sigma^{x}_{j} \sigma^{x}_{j+1}
    +\tilde{U}_{j} \sigma^{z}_{j} \sigma^{z}_{j+1}
    \right],
    \label{eq:Hspinherm}
\end{equation}
where $\tilde{U}_{j}= U_{j} - \i \delta_{j}$. As the XY spin model is exactly solvable~\cite{Schultz1964}, one can transform the above XZ spin model, which may not be exactly solvable, in Eq.~\eqref{eq:Hspinherm} into an XY model. For this purpose, we perform a $\pi/2$ spin rotation on all spin operators around the $x$ axis using $R =\exp[-\i \pi \sum_{j} \sigma^{x}_{j}/2]$~\cite{Miao2017}. The subsequent Hamiltonian yields
\begin{align}
    {\cal H} = \sum_{j}
    \left[ 
    -t_{j} \sigma^{x}_{j} \sigma^{x}_{j+1}
    +\tilde{U}_{j} \sigma^{y}_{j} \sigma^{y}_{j+1}
    \right].
    \label{eq:Hspinherm2}
\end{align}
To diagonalize this spin Hamiltonian, we introduce a second Jordan-Wigner transformation~\cite{Wang2017, Ezawa2017, Ding2021, Wada2021} given by
\begin{align}
    \sigma^{x}_{i} &=(f^{\dagger}_{i} + f_{i}) \exp \left[ \i \pi \sum_{i<j} f^{\dagger}_{i} f_{i} \right],\\
    \sigma^{y}_{i} &= -\i(f^{\dagger}_{i} - f_{i}) \exp \left[ \i \pi \sum_{i<j} f^{\dagger}_{i} f_{i} \right],\\
    \sigma^{z}_{i} &= 2 f^{\dagger}_{i} f_{i} - 1.
\end{align}
Using the above relations, we rewrite the XY spin model as a fermionic quadratic model, which reads
\begin{align}
    {\cal H} &= \sum_{j} - t_{j} \left[  f^{\dagger}_{j+1}f_{j} +f^{\dagger}_{j}  f_{j+1}
    + f^{\dagger}_{j} f^{\dagger}_{j+1}  + f_{j+1} f_{j} \right]\nonumber \\
    &+\sum_{j} \tilde{U}_{j} \left[ f^{\dagger}_{j+1}f_{j} +f^{\dagger}_{j}  f_{j+1}
    - f^{\dagger}_{j} f^{\dagger}_{j+1}  - f_{j+1} f_{j} \right].
    \label{eq:Hscf}
\end{align}

As the bulk-boundary correspondence is not violated in our non-Hermitian system, searching for the presence of zero modes can be done in the momentum space~($k$) using the Fourier transformation $f_{k} = \sum_{j} f_{j} \exp[\i k  j a ]$ with $a(=1)$ be the lattice constant.

\begin{figure}
    \centering
    \includegraphics[width=0.69\columnwidth]{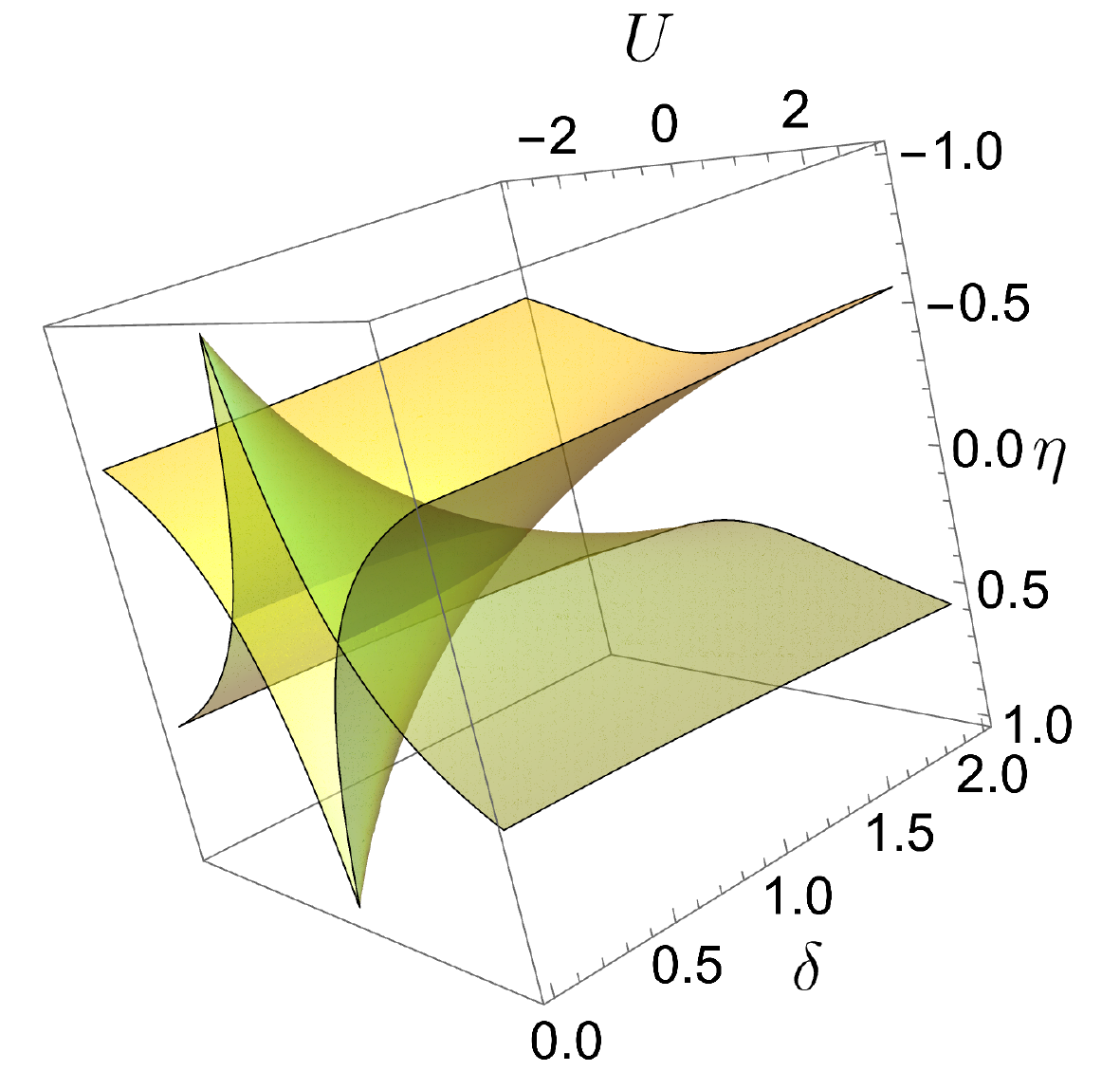}
    \caption{
    Phase boundaries of the effective Hamiltonian in Eq.~\eqref{eq:phasecond}. 
    \label{figsm:phasediag3D}
    }
\end{figure}

The quadratic Hamiltonian in the momentum space on a bipartite unit cell with sublattices $(A, B)$ casts
\begin{align}
    {\cal H} &=  \sum_{k} (-t +\tilde{U}) \left[  z_{k} f^{\dagger}_{k B} f_{kA} + z_{k}^{*} f^{\dagger}_{k A} f_{k B} \right]
    \nonumber \\
    &-\sum_{k} (t +\tilde{U}) \left[ 
    w_{k} f^{\dagger}_{-k A} f^{\dagger}_{kB} 
    + w_{k}^{*} f_{k B} f_{-k A} 
    \right]
    .
    \label{eq:Hscfk}
\end{align}
Here $\tilde{U}=U-\i \delta$ and $z_{k}$ and $w_{k}$ read
\begin{align} 
z_{k} &=(1+\eta) + e^{-\i k a} (1 - \eta) ,\\
w_{k} &= (1+ \eta) - e^{-\i k a} (1-\eta).
\end{align}
The non-Hermitian Hamiltonian in Eq.~\eqref{eq:Hscfk} can be written in a matrix form using the vector operator ${\cal C}_{k}^{\dagger}=(f^{\dagger}_{k A}, f^{\dagger}_{k B}, f_{-k A}, f_{-k B})$ as
\begin{align}
    {\cal H}&= \frac{1}{2} \sum_{k} {\cal C}^{\dagger}_{k}
    \begin{pmatrix}
    0 & a_{1} z_{k}^{*} & 0 &a_{2} w^{*}_{k} \\
    a_{1} z_{k} & 0 & -a_{2} w_{k} & 0 \\
    0 & -a_{2} w_{k}^{*} & 0 & - a_{1} z_{k}^{*} \\
    a_{2} w_{k} & 0 & -a_{1} z_{k} & 0
    \end{pmatrix}
    {\cal C}_{k}
    ,
    \label{eq:final_Hnuzero}
\end{align}
where $a_{1}  = -t +\tilde{U} $ and $a_{2} = -(t+\tilde{U})$. Expressing ${\cal H}= \frac{1}{2}  \sum_{k} \Lambda_{k} {\cal C}^{\dagger}_{k}  {\cal C}_{k}$, four eigenvalues $\Lambda_{k}$ then read
\begin{align}
    \frac{\Lambda_{k}^2}{4} &=
    \begin{cases}
    \tilde{U}^{2} (1 + \eta)^{2} + t^2 (1-\eta)^{2} -2 t \tilde{U}(1-\eta^{2}) \cos(ka)
    ,\\
    \tilde{U}^{2} (1 - \eta)^{2} + t^2 (1+\eta)^{2} -2 t \tilde{U}(1-\eta^{2}) \cos(ka)
,
    \end{cases}
    \label{eq:Lameig}
\end{align}
where $\tilde{U}^{2} = U^{2} -\delta^{2} - 2 \i \delta U$. 
The gap closure in $\Lambda_{k}$ occurs when we impose $\Re[\Lambda_{k}]=0$ resulting in
\begin{align}
    \frac{U}{t}=    \sqrt{\frac{\delta^2}{t^2} - \frac{(1 \pm \eta)^2}{(1 \mp \eta)^2 }},
    \label{eq:phasecond}
\end{align}
which is obtained at $k=\pm \pi/2$. We emphasize that due to the charge conjugation symmetry enforcing $\Re[\Lambda_{k}]=0$ ensures $\Im[\Lambda_{k}]=0$.

\begin{figure}
    \centering
    \includegraphics[width=0.99\columnwidth]{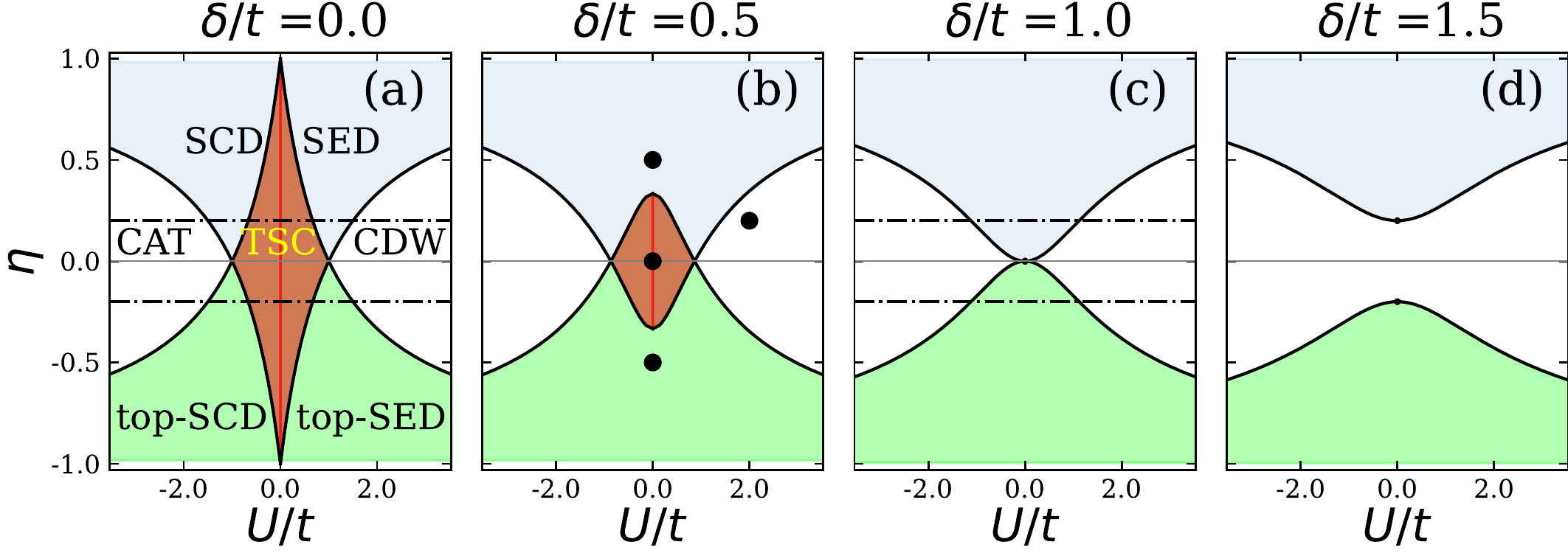}
    \caption{
    Phase diagrams of the non-Hermitian Hamiltonian in Eq.~\eqref{eq:phasecond} at $\mu=0$ on the $(U/t- \eta)$ plane for $\delta/t=$0.0~(a), 0.5~(b), 1.0~(c) and 1.5~(d). At $\delta=0$, the phase diagram consists of seven phases, including the topological superconducting~(TSC), (topological) superconducting dimer~((top-)SCD), (topological) single-electron dimer~((top-)SED), charge density wave~(CDW), and Schr\"odinger cat~(CAT) phases. Dashed dotted lines in (a) and (c) are at $\eta=\pm0.2$. The circle points in (b) are located at $(U/t, \eta) \in \{(0,0), (0,-0.5) (0,0.5), (2.0,0.2)\}$.
    \label{figsm:phasediag_nuzero}
    }
\end{figure}

While Eq.~\eqref{eq:phasecond} determines phase boundaries, we note that characterizing the properties of each phase cannot be addressed from the spectra of the Hamiltonian in Eq.~\eqref{eq:final_Hnuzero}. This is because the topological character of various phases may not be preserved after performing nonlocal (Wigner) transformations~\cite{McGinley2017, Ezawa2017}. Nevertheless, as non-Hermiticity does not violate the bulk-boundary correspondence, we can deduce some pieces of information regarding zero modes from effective noninteracting Hamiltonians. 

For this purpose, we start with presenting the phase boundaries in Eq.~\eqref{eq:phasecond} as a function of $(\eta, U, \delta)$ in Fig.~\ref{figsm:phasediag3D}. The two-dimensional intersection of this phase boundaries at $\delta=$ $0.0$~(a), $0.5$~(b), $1.0$~(c), and $1.5$~(d) is shown in Fig.~\ref{figsm:phasediag_nuzero}. The Hermitian phase diagram in Fig.~\ref{figsm:phasediag_nuzero}(a) comprises seven phases including the topological superconducting~(TSC), (topological) superconducting dimer~((top-)SCD), (topological) single-electron dimer~((top-)SED), charge density wave~(CDW), and Schr\"odinger cat~(CAT) phases~\cite{Ezawa2017}. The CAT phase is a superposition of two superconducting states~\cite{Miao2017, Ezawa2017}.

\begin{figure}
    \centering
    \includegraphics[width=0.98\columnwidth]{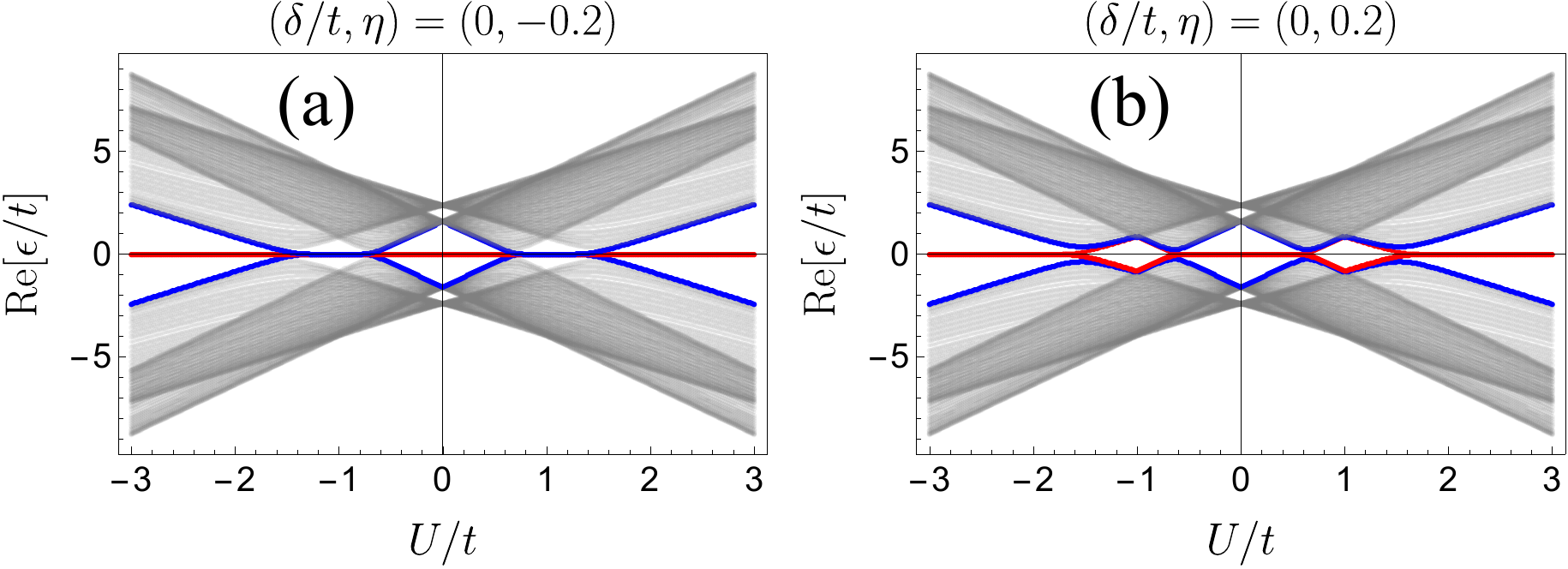}
    \caption{
    Spectra of the Hermitian Hamiltonian as a function of $U/t$ at $\mu=\delta=0$ and $\eta=-0.2$~(a) and $0.2$~(b). Two pairs of the smallest absolute values of eigenvalues are shown in red and blue.
    \label{figsm:spectra_EU_delta0}
    }
\end{figure}

To explore the zero modes in the phase diagram, we now look at the spectra of the system with an open boundary condition at $\mu=0$. We first consider the Hermitian limit with $\delta=0.0$ and at $\eta\pm 0.2$ along the cuts shown by dashed-dotted lines in Fig.~\ref{figsm:phasediag_nuzero}. The associated spectra for $\eta=-0.2(0.2)$ are shown in  Fig.~\ref{figsm:spectra_EU_delta0}.
Comparing two pairs of smallest absolute values of eigenvalues in (a, b), shown in red and blue, we realize that the CDW, CAT, and TSC phases are twofold degenerate.  
 The spectra also exhibit fourfold degeneracies in top-SCD and top-SED; see panel (a). However, these fourfold degeneracies are lifted for $\eta>0$ in the SCD and SED phases; see panel (b).

 \begin{figure}
    \centering
    \includegraphics[width=0.98\columnwidth]{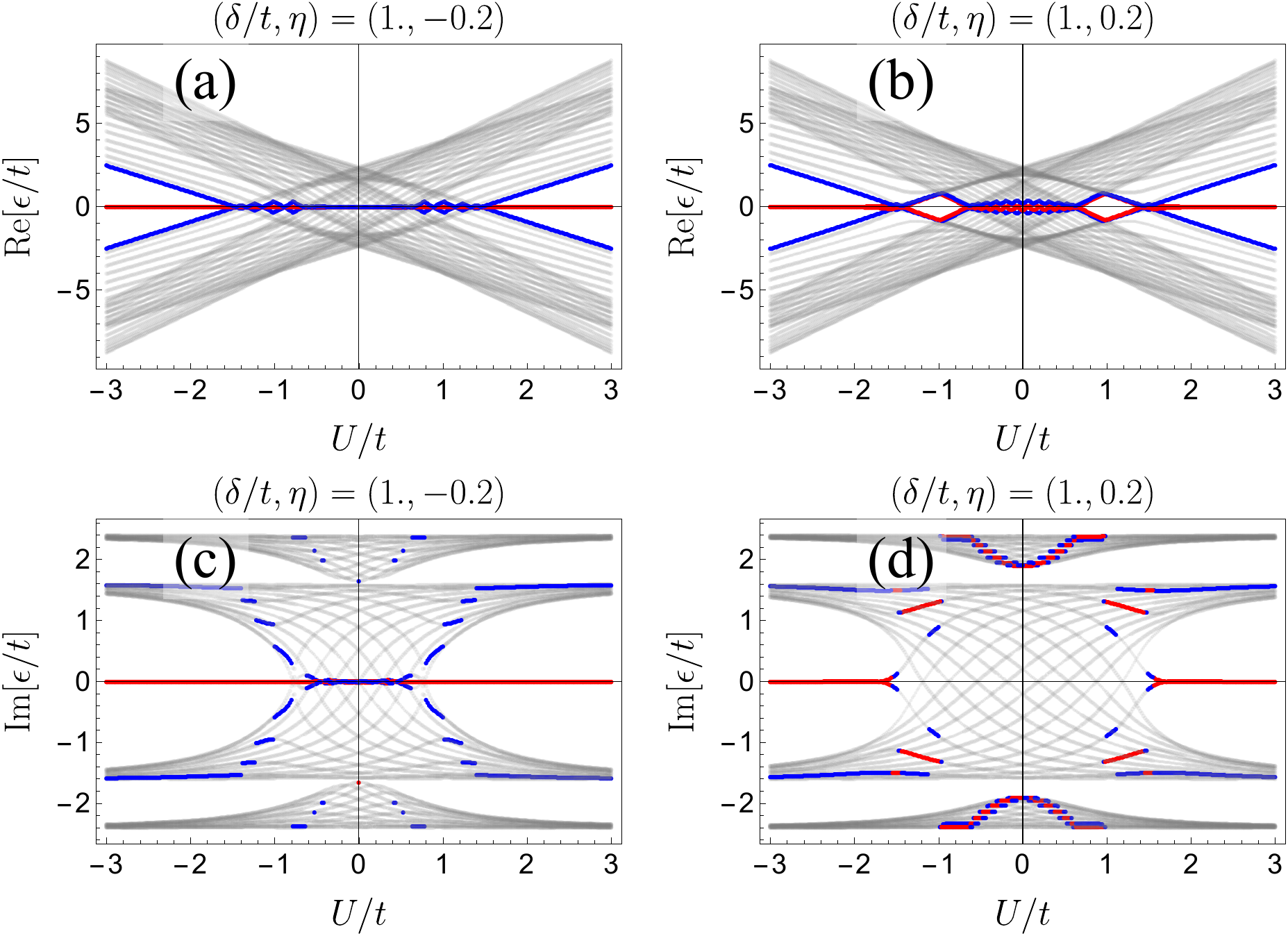}
    \caption{
    Real(a,b) and imaginary~(c,d) parts of the spectra as a function of $U/t$ at $\mu=0$, $\delta=1.0$ and $\eta=-0.2$(a,c) and $0.2$(b,d). Two pairs of the smallest absolute values of eigenvalues are shown in red and blue.
    \label{figsm:spectra_EU_delta1}
    }
\end{figure}

Further looking at Fig.~\ref{figsm:phasediag_nuzero} reveals that the boundaries of the TSC phase shrink toward diminishing this phase as $\delta$ increases. Reducing the regions with the TSC phase results in lifting the separation between pairs of phases, namely (top-SCD, top-SED), (SCD, SED), and (CAT, CDW). As the order of degeneracies in both components of these pairs are identical, we still detect fourfold, twofold, and no degeneracies, respectively, in the green, white, and blue regions in all panels of Fig.~\ref{figsm:phasediag_nuzero}. This can be seen in the spectra of the system at $\delta=1.0$ along cuts at $\eta=\pm0.2$ in Fig.~\ref{figsm:spectra_EU_delta1}. Note that the real and imaginary parts of eigenvalues are zero at degenerate points due to the particle-hole symmetry in the system.

 \begin{figure}
    \centering
        \includegraphics[width=0.98\columnwidth]{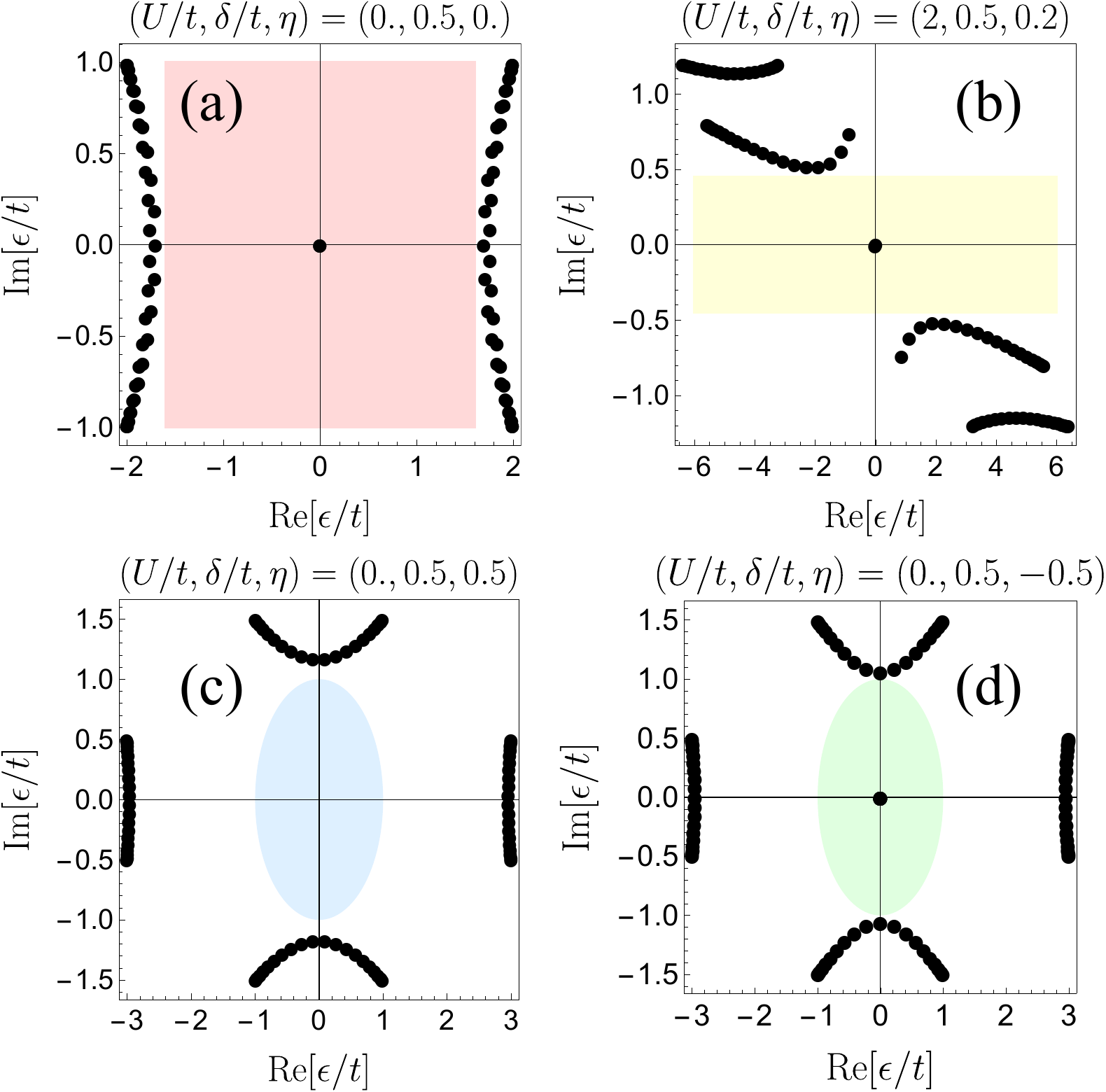}
    \caption{
    Spectra at $\mu=0$, $\delta=0.5$ and $(U/t, \eta)=(0,0)$(a), $(2,0.2)$(b), $(0,-0.5)$(c), and $(0,0.5)$(d). The real~(imaginary) line gaps are shown in red~(yellow). The point gap is shown by a circle with a unit radius.
    \label{figsm:spectra_ReIm_delta0.5}
    }
\end{figure}

 Aside from $2n$-fold degeneracies with $n\in\{0,1,2\}$, each non-Hermitian phase in Fig.~\ref{figsm:phasediag_nuzero} exhibits a particular non-Hermitian gap in their spectrum. We demonstrate this point by plotting the complex spectra of systems belonging to the brown, blue, lime, and white regions, shown by circle points in Fig.~\ref{figsm:phasediag_nuzero}(b). In the TSC phases, the spectrum displays a real line gap shown by a red rectangle in Fig.~\ref{figsm:spectra_ReIm_delta0.5}~(a). The spectra exhibit the imaginary line gap within the CDW and CAT phases as exemplified in Fig~\ref{figsm:spectra_ReIm_delta0.5}(b). The gap becomes the point gap both in phases with fourfold or no degeneracies, e.g., in the top-SED shown in Fig~\ref{figsm:spectra_ReIm_delta0.5}(c) and in the SED phase presented in Fig.~\ref{figsm:spectra_ReIm_delta0.5}(d).

\section{Topological modes in the thermodynamic limit}
\begin{figure}
    \centering
    \includegraphics[width=\columnwidth]{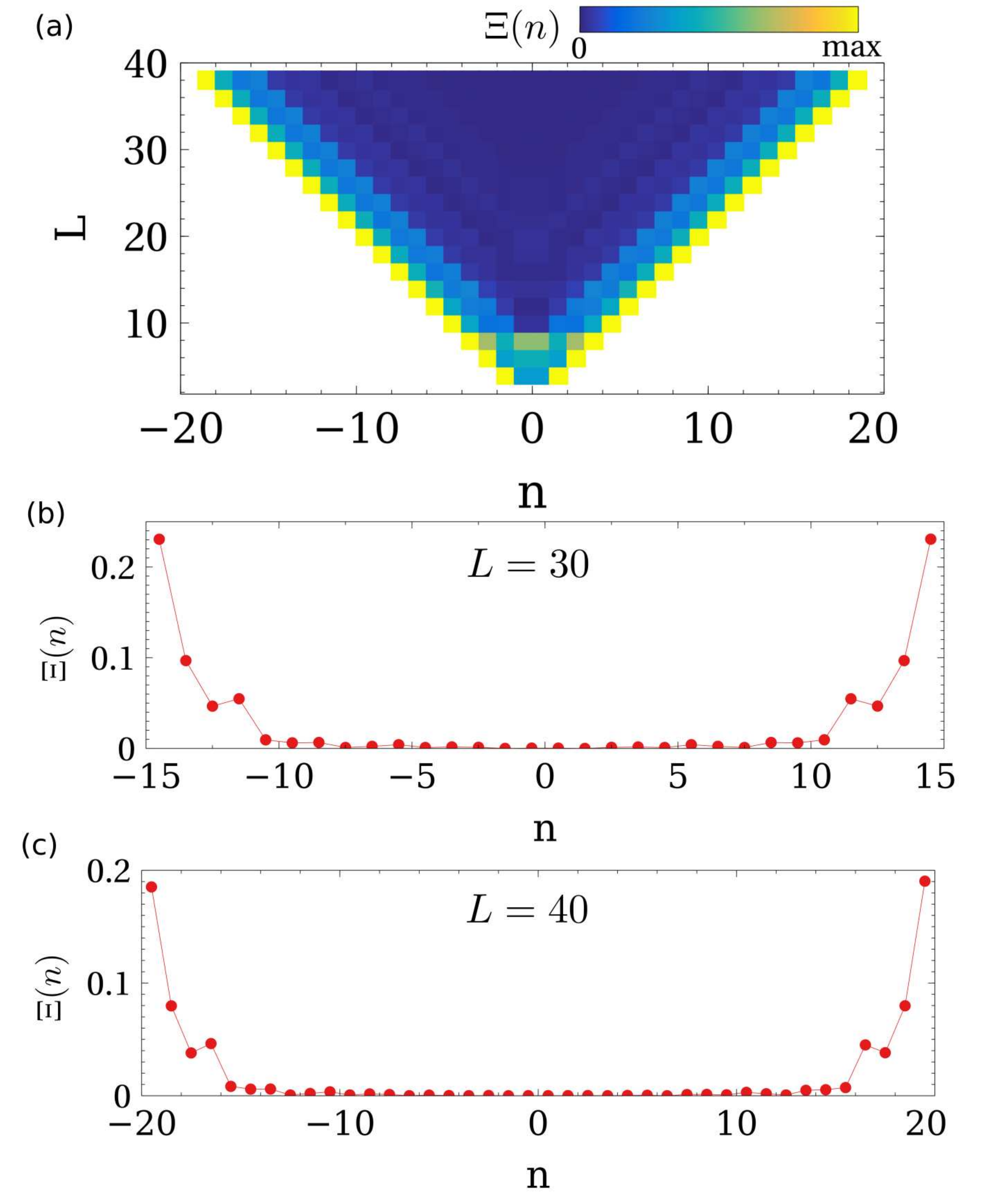}
    \caption{ Spatially resolved local correlators, showing the emergence of edge modes.
    Panel (a) shows the scaling with the length of the system, highlighting the emergence of decoupled modes
    for large lengths. Panels (b,c) show the spatially resolved correlators for two specific lengths,
    showing the localization of the excitations on the edges. We set  $U=0.5t$, $\Delta=t$, $\mu=0.25t$, $\delta=0.5t$ and $\eta=0$.}
    \label{fig:mps}
    
\end{figure}

In the previous sections, we have focused on relatively small-length systems that can be solved with exact
diagonalization. In this section, we show that the edge modes associated with topological degeneracies
evolve into fully decoupled modes in the thermodynamic limit. In order to surpass the length limitations
of exact diagonalization, in this section, we solve the interacting quantum many-body model
using a non-Hermitian tensor network formalism~\cite{dmrgpy, ITensor, Fishman2022}, targeting the ground state and lowest excited states~\cite{Hyart2022, Chen2022a}. We show in Fig.~\ref{fig:mps} the spatially resolved correlator
computed with the non-Hermitian tensor network formalism. It is observed that as the system becomes larger,
the local edge modes become more decoupled (Fig.~\ref{fig:mps}(a)), giving rise to fully decoupled modes
for large systems; see panels (b) and (c) in Fig.~\ref{fig:mps}.

\begin{figure}[t!]
    \centering
    \includegraphics[width=\columnwidth]{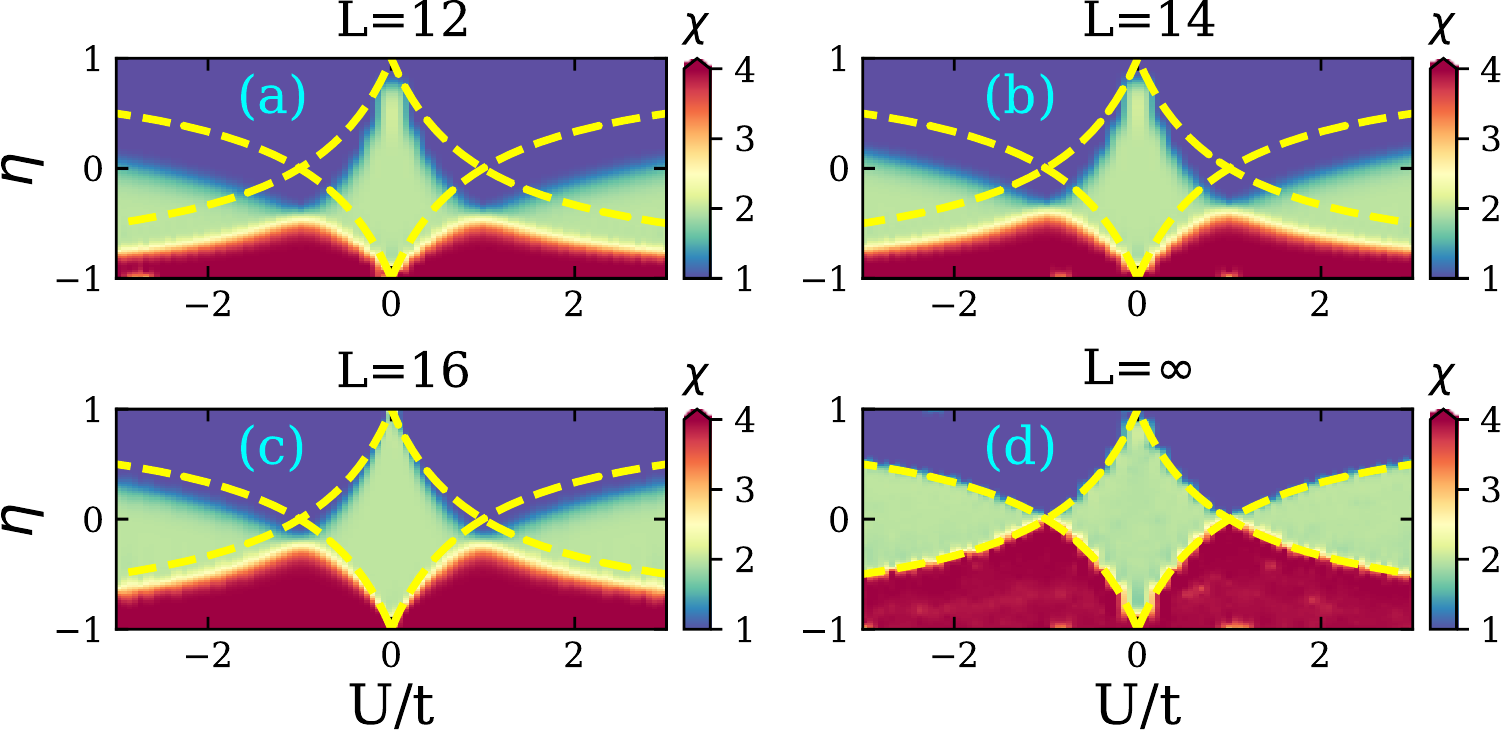}
    \caption{Degeneracy of the non-Hermitian model for different system sizes, $L=12$~(a), $L=14$~(b)
    and $L=16$~(c), and extrapolation in the thermodynamic limit (d). It is observed that while for finite systems, finite size effects affect the inferred boundaries between different phases, the extrapolation scheme allows to the recovery of the correct phase diagram. The yellow dashed lines display the exact phase boundaries. We took $\Delta=t$, $\delta=\mu=0$, and $\lambda=0.01t$.
    \label{fig:extrapolation}
    }
\end{figure}

\section{Finite size effects and extrapolation to the thermodynamic limit}

In the main text, we have focused on studying the case of finite systems numerically. Due to finite
size effects, the degeneracy of the ground state is lifted due to the coupling of the topological excitations.
In order to extract the topological degeneracy in the thermodynamic limit, a finite size scaling of the energy splittings can be performed. In this section, we show that with a size scaling, we can recover the exact analytic results in the interacting limit from exact numerical calculations in finite systems. The finite size scaling relies on taking a scaling for the excited state energies as $\epsilon_\alpha(L) = a_\alpha + b_\alpha/L$, so that in the thermodynamic limit $\epsilon_\alpha(L=\infty) = a_\alpha$. The excited state energies are taken as $\epsilon_\alpha = |E_\alpha - E_0|$, where $E_\alpha$ is the complex eigenenergy of the many-body Hamiltonian and $E_0$ the ground state energy. By computing the energies for a set of finite-size systems, the coefficients $a,b$ can be extracted, and the energies in the thermodynamic limit are obtained. With the extracted energy differences, the degeneracy of the ground state can be computed as
$\chi = \sum_\alpha e^{-\lambda \epsilon_\alpha}$ with $\lambda$ being the energy smearing.

We present in Fig.~\ref{fig:extrapolation} how this methodology allows obtaining the degeneracies in the thermodynamic limit. It is observed that for finite-size systems, the phase boundaries are substantially shifted and are size dependent; see panels (a), (b), and (c) in Fig.~\ref{fig:extrapolation}). Using the extrapolation scheme noted above, the correct phase boundaries known from the analytic solution are recovered, as shown in Fig.~\ref{fig:extrapolation}(d).

\begin{figure}[t!]
    \centering
    \includegraphics[width=\columnwidth]{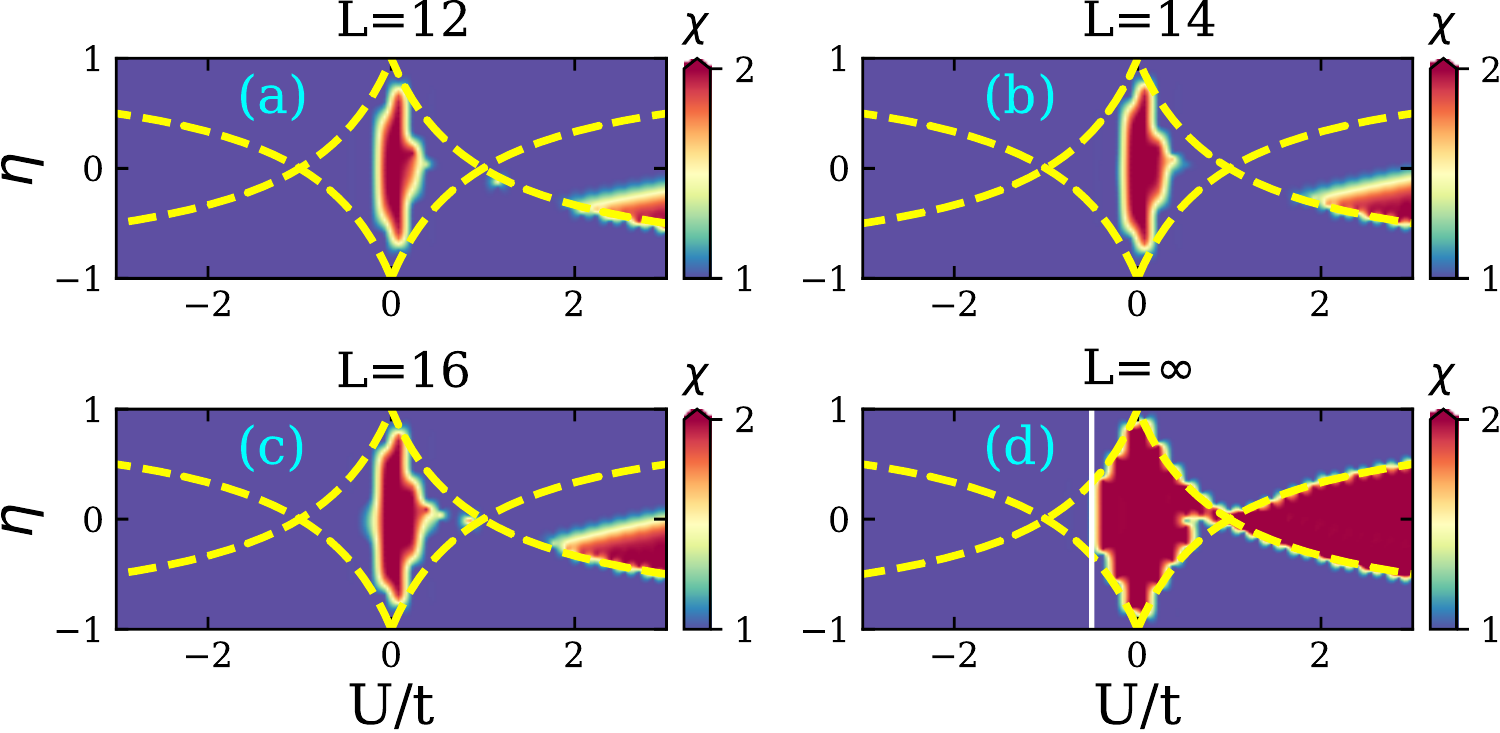}
    \caption{Degeneracy of the non-Hermitian model for different system sizes, $L=12$~(a), $L=14$~(b)
    and $L=16$~(c), and extrapolation in the thermodynamic limit (d) with $\Delta=t$, $\delta=0$, $\mu=0.25t$, and $\lambda=0.01t$. The vertical white line in (d) is at $U=-0.5t$.
    \label{fig:extrapolation_d0mu025}
    }
\end{figure}

\begin{figure}[t!]
    \centering
    \includegraphics[width=\columnwidth]{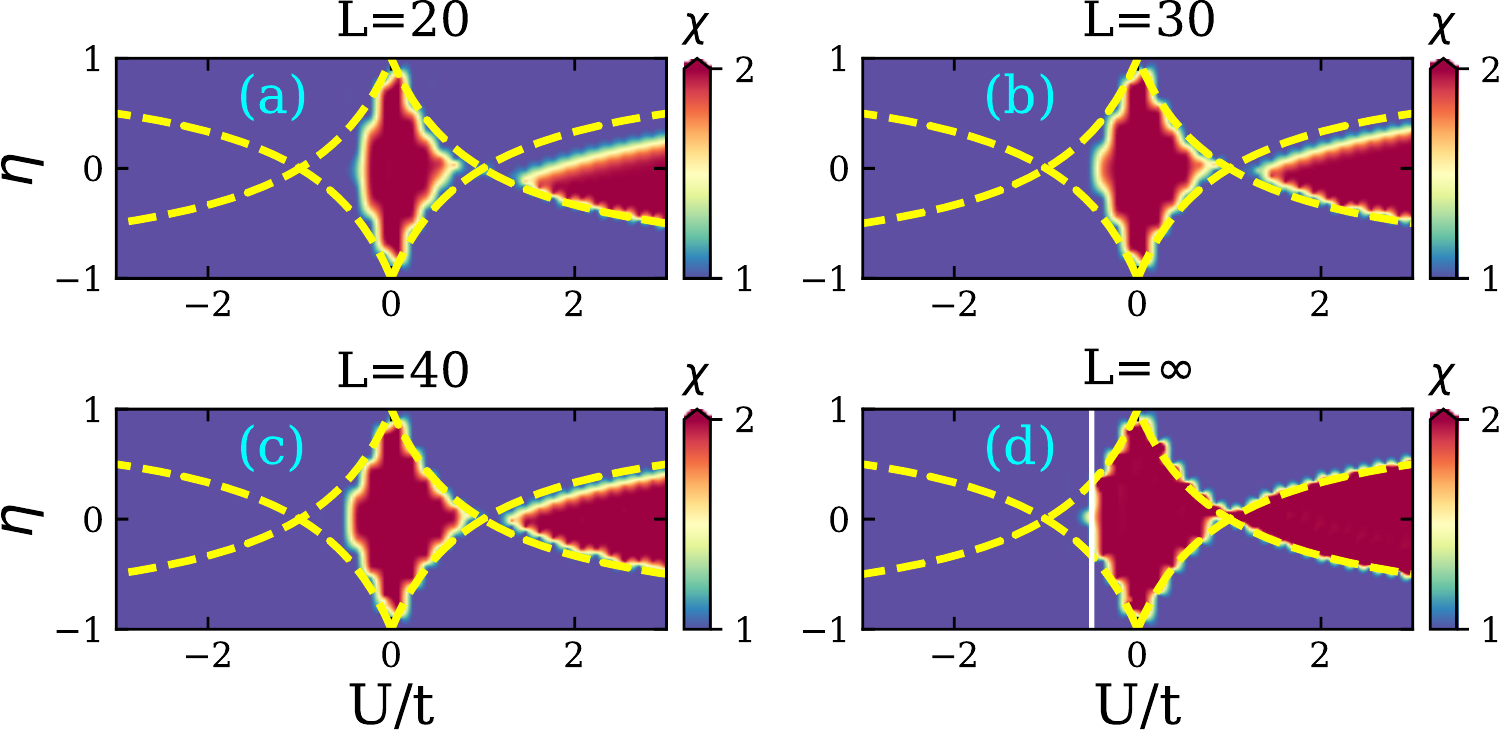}
    \caption{Degeneracy of the non-Hermitian model obtained from the DMRG calculations for different system sizes, $L=20$~(a), $L=30$~(b)
    and $L=40$~(c), and extrapolation in the thermodynamic limit (d) with $\Delta=t$, $\delta=0$, $\mu=0.25t$ and $\lambda=0.01t$. The vertical white line in (d) is at $U=-0.5t$.
    \label{fig:extrapolation_d0mu025_dmrg}
    }
\end{figure}

We further perform a similar extrapolation scheme to the Hermitian phase diagram at $\mu=0.25t$. Here we employ the exact diagonalization methods in Fig.~\ref{fig:extrapolation_d0mu025} and the tensor-network formalism in Fig.~\ref{fig:extrapolation_d0mu025_dmrg}. As it is evident, increasing the length of the chain in panels (a)-(c) results in improving the boundaries of twofold degenerate topological phases. In the thermodynamic limit, the extrapolated phase diagrams, presented in Fig.~\ref{fig:extrapolation_d0mu025}(d) and Fig.~\ref{fig:extrapolation_d0mu025_dmrg}(d), recover the exact phase boundaries at $\mu=0$ and $U\geq -0.5t$. We note that the phase boundary at $\eta=0$ is located at $U=-0.5t$~(vertical white line), in agreement with the previous Hermitian calculations~\cite{Mahyaeh2020}.

\bibliography{int_dim_kitaev}

\begin{thebibliography}{100}%
\makeatletter
\providecommand \@ifxundefined [1]{%
 \@ifx{#1\undefined}
}%
\providecommand \@ifnum [1]{%
 \ifnum #1\expandafter \@firstoftwo
 \else \expandafter \@secondoftwo
 \fi
}%
\providecommand \@ifx [1]{%
 \ifx #1\expandafter \@firstoftwo
 \else \expandafter \@secondoftwo
 \fi
}%
\providecommand \natexlab [1]{#1}%
\providecommand \enquote  [1]{``#1''}%
\providecommand \bibnamefont  [1]{#1}%
\providecommand \bibfnamefont [1]{#1}%
\providecommand \citenamefont [1]{#1}%
\providecommand \href@noop [0]{\@secondoftwo}%
\providecommand \href [0]{\begingroup \@sanitize@url \@href}%
\providecommand \@href[1]{\@@startlink{#1}\@@href}%
\providecommand \@@href[1]{\endgroup#1\@@endlink}%
\providecommand \@sanitize@url [0]{\catcode `\\12\catcode `\$12\catcode
  `\&12\catcode `\#12\catcode `\^12\catcode `\_12\catcode `\%12\relax}%
\providecommand \@@startlink[1]{}%
\providecommand \@@endlink[0]{}%
\providecommand \url  [0]{\begingroup\@sanitize@url \@url }%
\providecommand \@url [1]{\endgroup\@href {#1}{\urlprefix }}%
\providecommand \urlprefix  [0]{URL }%
\providecommand \Eprint [0]{\href }%
\providecommand \doibase [0]{http://dx.doi.org/}%
\providecommand \selectlanguage [0]{\@gobble}%
\providecommand \bibinfo  [0]{\@secondoftwo}%
\providecommand \bibfield  [0]{\@secondoftwo}%
\providecommand \translation [1]{[#1]}%
\providecommand \BibitemOpen [0]{}%
\providecommand \bibitemStop [0]{}%
\providecommand \bibitemNoStop [0]{.\EOS\space}%
\providecommand \EOS [0]{\spacefactor3000\relax}%
\providecommand \BibitemShut  [1]{\csname bibitem#1\endcsname}%
\let\auto@bib@innerbib\@empty
\bibitem [{\citenamefont {Koma}\ and\ \citenamefont {Tasaki}(1994)}]{Koma1994}%
  \BibitemOpen
  \bibfield  {author} {\bibinfo {author} {\bibfnamefont {Tohru}\ \bibnamefont
  {Koma}}\ and\ \bibinfo {author} {\bibfnamefont {Hal}\ \bibnamefont
  {Tasaki}},\ }\bibfield  {title} {\enquote {\bibinfo {title} {Symmetry
  breaking and finite-size effects in quantum many-body systems},}\ }\href
  {\doibase 10.1007/bf02188685} {\bibfield  {journal} {\bibinfo  {journal}
  {Journal of Statistical Physics}\ }\textbf {\bibinfo {volume} {76}},\
  \bibinfo {pages} {745--803} (\bibinfo {year} {1994})}\BibitemShut {NoStop}%
\bibitem [{\citenamefont {Belyaev}(2006)}]{Belyaev2006}%
  \BibitemOpen
  \bibfield  {author} {\bibinfo {author} {\bibfnamefont {Spartak~T.}\
  \bibnamefont {Belyaev}},\ }\bibfield  {title} {\enquote {\bibinfo {title}
  {Many-body physics and spontaneous symmetry breaking},}\ }\href {\doibase
  10.1142/S0217979206035059} {\bibfield  {journal} {\bibinfo  {journal}
  {International Journal of Modern Physics B}\ }\textbf {\bibinfo {volume}
  {20}},\ \bibinfo {pages} {2579--2590} (\bibinfo {year} {2006})},\ \Eprint
  {http://arxiv.org/abs/https://doi.org/10.1142/S0217979206035059}
  {https://doi.org/10.1142/S0217979206035059} \BibitemShut {NoStop}%
\bibitem [{\citenamefont {Brauner}(2010)}]{Brauner2010}%
  \BibitemOpen
  \bibfield  {author} {\bibinfo {author} {\bibfnamefont {Tomas}\ \bibnamefont
  {Brauner}},\ }\bibfield  {title} {\enquote {\bibinfo {title} {Spontaneous
  symmetry breaking and nambu–goldstone bosons in quantum many-body
  systems},}\ }\href {\doibase 10.3390/sym2020609} {\bibfield  {journal}
  {\bibinfo  {journal} {Symmetry}\ }\textbf {\bibinfo {volume} {2}},\ \bibinfo
  {pages} {609--657} (\bibinfo {year} {2010})}\BibitemShut {NoStop}%
\bibitem [{\citenamefont {Dong}\ \emph {et~al.}(2017)\citenamefont {Dong},
  \citenamefont {Fang},\ and\ \citenamefont {Sun}}]{Dong2017}%
  \BibitemOpen
  \bibfield  {author} {\bibinfo {author} {\bibfnamefont {Guo-Hui}\ \bibnamefont
  {Dong}}, \bibinfo {author} {\bibfnamefont {Yi-Nan}\ \bibnamefont {Fang}}, \
  and\ \bibinfo {author} {\bibfnamefont {Chang-Pu}\ \bibnamefont {Sun}},\
  }\bibfield  {title} {\enquote {\bibinfo {title} {Quantifying spontaneously
  symmetry breaking of quantum many-body systems*},}\ }\href {\doibase
  10.1088/0253-6102/68/4/405} {\bibfield  {journal} {\bibinfo  {journal}
  {Communications in Theoretical Physics}\ }\textbf {\bibinfo {volume} {68}},\
  \bibinfo {pages} {405} (\bibinfo {year} {2017})}\BibitemShut {NoStop}%
\bibitem [{\citenamefont {Santos}\ \emph {et~al.}(2016)\citenamefont {Santos},
  \citenamefont {T\'avora},\ and\ \citenamefont {P\'erez-Bernal}}]{Santos2016}%
  \BibitemOpen
  \bibfield  {author} {\bibinfo {author} {\bibfnamefont {Lea~F.}\ \bibnamefont
  {Santos}}, \bibinfo {author} {\bibfnamefont {Marco}\ \bibnamefont
  {T\'avora}}, \ and\ \bibinfo {author} {\bibfnamefont {Francisco}\
  \bibnamefont {P\'erez-Bernal}},\ }\bibfield  {title} {\enquote {\bibinfo
  {title} {Excited-state quantum phase transitions in many-body systems with
  infinite-range interaction: Localization, dynamics, and bifurcation},}\
  }\href {\doibase 10.1103/PhysRevA.94.012113} {\bibfield  {journal} {\bibinfo
  {journal} {Phys. Rev. A}\ }\textbf {\bibinfo {volume} {94}},\ \bibinfo
  {pages} {012113} (\bibinfo {year} {2016})}\BibitemShut {NoStop}%
\bibitem [{\citenamefont {Heyl}(2018)}]{Heyl2018}%
  \BibitemOpen
  \bibfield  {author} {\bibinfo {author} {\bibfnamefont {Markus}\ \bibnamefont
  {Heyl}},\ }\bibfield  {title} {\enquote {\bibinfo {title} {Dynamical quantum
  phase transitions: a review},}\ }\href {\doibase 10.1088/1361-6633/aaaf9a}
  {\bibfield  {journal} {\bibinfo  {journal} {Reports on Progress in Physics}\
  }\textbf {\bibinfo {volume} {81}},\ \bibinfo {pages} {054001} (\bibinfo
  {year} {2018})}\BibitemShut {NoStop}%
\bibitem [{\citenamefont {Carollo}\ \emph {et~al.}(2020)\citenamefont
  {Carollo}, \citenamefont {Valenti},\ and\ \citenamefont
  {Spagnolo}}]{Carollo2020}%
  \BibitemOpen
  \bibfield  {author} {\bibinfo {author} {\bibfnamefont {Angelo}\ \bibnamefont
  {Carollo}}, \bibinfo {author} {\bibfnamefont {Davide}\ \bibnamefont
  {Valenti}}, \ and\ \bibinfo {author} {\bibfnamefont {Bernardo}\ \bibnamefont
  {Spagnolo}},\ }\bibfield  {title} {\enquote {\bibinfo {title} {Geometry of
  quantum phase transitions},}\ }\href {\doibase 10.1016/j.physrep.2019.11.002}
  {\bibfield  {journal} {\bibinfo  {journal} {Physics Reports}\ }\textbf
  {\bibinfo {volume} {838}},\ \bibinfo {pages} {1--72} (\bibinfo {year}
  {2020})}\BibitemShut {NoStop}%
\bibitem [{\citenamefont {Serwatka}\ \emph {et~al.}(2023)\citenamefont
  {Serwatka}, \citenamefont {Melko}, \citenamefont {Burkov},\ and\
  \citenamefont {Roy}}]{Serwatka2023}%
  \BibitemOpen
  \bibfield  {author} {\bibinfo {author} {\bibfnamefont {T.}~\bibnamefont
  {Serwatka}}, \bibinfo {author} {\bibfnamefont {R.~G.}\ \bibnamefont {Melko}},
  \bibinfo {author} {\bibfnamefont {A.}~\bibnamefont {Burkov}}, \ and\ \bibinfo
  {author} {\bibfnamefont {P.-N.}\ \bibnamefont {Roy}},\ }\bibfield  {title}
  {\enquote {\bibinfo {title} {Quantum phase transition in the one-dimensional
  water chain},}\ }\href {\doibase 10.1103/PhysRevLett.130.026201} {\bibfield
  {journal} {\bibinfo  {journal} {Phys. Rev. Lett.}\ }\textbf {\bibinfo
  {volume} {130}},\ \bibinfo {pages} {026201} (\bibinfo {year}
  {2023})}\BibitemShut {NoStop}%
\bibitem [{\citenamefont {Zhang}\ \emph {et~al.}(2020)\citenamefont {Zhang},
  \citenamefont {Zhang}, \citenamefont {Li}, \citenamefont {Wang},\ and\
  \citenamefont {Zhu}}]{Zhang2020}%
  \BibitemOpen
  \bibfield  {author} {\bibinfo {author} {\bibfnamefont {Guo-Qing}\
  \bibnamefont {Zhang}}, \bibinfo {author} {\bibfnamefont {Dan-Wei}\
  \bibnamefont {Zhang}}, \bibinfo {author} {\bibfnamefont {Zhi}\ \bibnamefont
  {Li}}, \bibinfo {author} {\bibfnamefont {Z.~D.}\ \bibnamefont {Wang}}, \ and\
  \bibinfo {author} {\bibfnamefont {Shi-Liang}\ \bibnamefont {Zhu}},\
  }\bibfield  {title} {\enquote {\bibinfo {title} {Statistically related
  many-body localization in the one-dimensional anyon hubbard model},}\ }\href
  {\doibase 10.1103/physrevb.102.054204} {\bibfield  {journal} {\bibinfo
  {journal} {Physical Review B}\ }\textbf {\bibinfo {volume} {102}} (\bibinfo
  {year} {2020}),\ 10.1103/physrevb.102.054204}\BibitemShut {NoStop}%
\bibitem [{\citenamefont {Hashisaka}\ \emph {et~al.}(2021)\citenamefont
  {Hashisaka}, \citenamefont {Jonckheere}, \citenamefont {Akiho}, \citenamefont
  {Sasaki}, \citenamefont {Rech}, \citenamefont {Martin},\ and\ \citenamefont
  {Muraki}}]{Hashisaka2021}%
  \BibitemOpen
  \bibfield  {author} {\bibinfo {author} {\bibfnamefont {M.}~\bibnamefont
  {Hashisaka}}, \bibinfo {author} {\bibfnamefont {T.}~\bibnamefont
  {Jonckheere}}, \bibinfo {author} {\bibfnamefont {T.}~\bibnamefont {Akiho}},
  \bibinfo {author} {\bibfnamefont {S.}~\bibnamefont {Sasaki}}, \bibinfo
  {author} {\bibfnamefont {J.}~\bibnamefont {Rech}}, \bibinfo {author}
  {\bibfnamefont {T.}~\bibnamefont {Martin}}, \ and\ \bibinfo {author}
  {\bibfnamefont {K.}~\bibnamefont {Muraki}},\ }\bibfield  {title} {\enquote
  {\bibinfo {title} {Andreev reflection of fractional quantum hall
  quasiparticles},}\ }\href {\doibase 10.1038/s41467-021-23160-6} {\bibfield
  {journal} {\bibinfo  {journal} {Nature Communications}\ }\textbf {\bibinfo
  {volume} {12}} (\bibinfo {year} {2021}),\
  10.1038/s41467-021-23160-6}\BibitemShut {NoStop}%
\bibitem [{\citenamefont {Kaskela}\ and\ \citenamefont
  {Lado}(2021)}]{Kaskela2021}%
  \BibitemOpen
  \bibfield  {author} {\bibinfo {author} {\bibfnamefont {Vilja}\ \bibnamefont
  {Kaskela}}\ and\ \bibinfo {author} {\bibfnamefont {J.~L.}\ \bibnamefont
  {Lado}},\ }\bibfield  {title} {\enquote {\bibinfo {title} {Dynamical
  topological excitations in parafermion chains},}\ }\href {\doibase
  10.1103/PhysRevResearch.3.013095} {\bibfield  {journal} {\bibinfo  {journal}
  {Phys. Rev. Res.}\ }\textbf {\bibinfo {volume} {3}},\ \bibinfo {pages}
  {013095} (\bibinfo {year} {2021})}\BibitemShut {NoStop}%
\bibitem [{\citenamefont {Wouters}\ \emph {et~al.}(2022)\citenamefont
  {Wouters}, \citenamefont {Hassler}, \citenamefont {Katsura},\ and\
  \citenamefont {Schuricht}}]{Wouters2022}%
  \BibitemOpen
  \bibfield  {author} {\bibinfo {author} {\bibfnamefont {Jurriaan}\
  \bibnamefont {Wouters}}, \bibinfo {author} {\bibfnamefont {Fabian}\
  \bibnamefont {Hassler}}, \bibinfo {author} {\bibfnamefont {Hosho}\
  \bibnamefont {Katsura}}, \ and\ \bibinfo {author} {\bibfnamefont {Dirk}\
  \bibnamefont {Schuricht}},\ }\bibfield  {title} {\enquote {\bibinfo {title}
  {{Phase diagram of an extended parafermion chain}},}\ }\href {\doibase
  10.21468/SciPostPhysCore.5.1.008} {\bibfield  {journal} {\bibinfo  {journal}
  {SciPost Phys. Core}\ }\textbf {\bibinfo {volume} {5}},\ \bibinfo {pages}
  {008} (\bibinfo {year} {2022})}\BibitemShut {NoStop}%
\bibitem [{\citenamefont {Gangadharaiah}\ \emph {et~al.}(2011)\citenamefont
  {Gangadharaiah}, \citenamefont {Braunecker}, \citenamefont {Simon},\ and\
  \citenamefont {Loss}}]{Gangadharaiah2011}%
  \BibitemOpen
  \bibfield  {author} {\bibinfo {author} {\bibfnamefont {Suhas}\ \bibnamefont
  {Gangadharaiah}}, \bibinfo {author} {\bibfnamefont {Bernd}\ \bibnamefont
  {Braunecker}}, \bibinfo {author} {\bibfnamefont {Pascal}\ \bibnamefont
  {Simon}}, \ and\ \bibinfo {author} {\bibfnamefont {Daniel}\ \bibnamefont
  {Loss}},\ }\bibfield  {title} {\enquote {\bibinfo {title} {Majorana edge
  states in interacting one-dimensional systems},}\ }\href {\doibase
  10.1103/PhysRevLett.107.036801} {\bibfield  {journal} {\bibinfo  {journal}
  {Phys. Rev. Lett.}\ }\textbf {\bibinfo {volume} {107}},\ \bibinfo {pages}
  {036801} (\bibinfo {year} {2011})}\BibitemShut {NoStop}%
\bibitem [{\citenamefont {Katsura}\ \emph {et~al.}(2015)\citenamefont
  {Katsura}, \citenamefont {Schuricht},\ and\ \citenamefont
  {Takahashi}}]{Katsura2015}%
  \BibitemOpen
  \bibfield  {author} {\bibinfo {author} {\bibfnamefont {Hosho}\ \bibnamefont
  {Katsura}}, \bibinfo {author} {\bibfnamefont {Dirk}\ \bibnamefont
  {Schuricht}}, \ and\ \bibinfo {author} {\bibfnamefont {Masahiro}\
  \bibnamefont {Takahashi}},\ }\bibfield  {title} {\enquote {\bibinfo {title}
  {Exact ground states and topological order in interacting kitaev/majorana
  chains},}\ }\href {\doibase 10.1103/PhysRevB.92.115137} {\bibfield  {journal}
  {\bibinfo  {journal} {Phys. Rev. B}\ }\textbf {\bibinfo {volume} {92}},\
  \bibinfo {pages} {115137} (\bibinfo {year} {2015})}\BibitemShut {NoStop}%
\bibitem [{\citenamefont {Ezawa}(2017)}]{Ezawa2017}%
  \BibitemOpen
  \bibfield  {author} {\bibinfo {author} {\bibfnamefont {Motohiko}\
  \bibnamefont {Ezawa}},\ }\bibfield  {title} {\enquote {\bibinfo {title}
  {{Exact solutions and topological phase diagram in interacting dimerized
  Kitaev topological superconductors}},}\ }\href {\doibase
  10.1103/PhysRevB.96.121105} {\bibfield  {journal} {\bibinfo  {journal}
  {Physical Review B}\ }\textbf {\bibinfo {volume} {96}},\ \bibinfo {pages}
  {1--5} (\bibinfo {year} {2017})},\ \Eprint {http://arxiv.org/abs/1707.03983}
  {arXiv:1707.03983} \BibitemShut {NoStop}%
\bibitem [{\citenamefont {Wang}\ \emph {et~al.}(2017)\citenamefont {Wang},
  \citenamefont {Miao}, \citenamefont {Jin},\ and\ \citenamefont
  {Chen}}]{Wang2017}%
  \BibitemOpen
  \bibfield  {author} {\bibinfo {author} {\bibfnamefont {Yucheng}\ \bibnamefont
  {Wang}}, \bibinfo {author} {\bibfnamefont {Jian~Jian}\ \bibnamefont {Miao}},
  \bibinfo {author} {\bibfnamefont {Hui~Ke}\ \bibnamefont {Jin}}, \ and\
  \bibinfo {author} {\bibfnamefont {Shu}\ \bibnamefont {Chen}},\ }\bibfield
  {title} {\enquote {\bibinfo {title} {{Characterization of topological phases
  of dimerized Kitaev chain via edge correlation functions}},}\ }\href
  {\doibase 10.1103/PhysRevB.96.205428} {\bibfield  {journal} {\bibinfo
  {journal} {Physical Review B}\ }\textbf {\bibinfo {volume} {96}} (\bibinfo
  {year} {2017}),\ 10.1103/PhysRevB.96.205428},\ \Eprint
  {http://arxiv.org/abs/1708.03891} {arXiv:1708.03891} \BibitemShut {NoStop}%
\bibitem [{\citenamefont {Miao}\ \emph {et~al.}(2017)\citenamefont {Miao},
  \citenamefont {Jin}, \citenamefont {Zhang},\ and\ \citenamefont
  {Zhou}}]{Miao2017}%
  \BibitemOpen
  \bibfield  {author} {\bibinfo {author} {\bibfnamefont {Jian-Jian}\
  \bibnamefont {Miao}}, \bibinfo {author} {\bibfnamefont {Hui-Ke}\ \bibnamefont
  {Jin}}, \bibinfo {author} {\bibfnamefont {Fu-Chun}\ \bibnamefont {Zhang}}, \
  and\ \bibinfo {author} {\bibfnamefont {Yi}~\bibnamefont {Zhou}},\ }\bibfield
  {title} {\enquote {\bibinfo {title} {Exact solution for the interacting
  kitaev chain at the symmetric point},}\ }\href {\doibase
  10.1103/PhysRevLett.118.267701} {\bibfield  {journal} {\bibinfo  {journal}
  {Phys. Rev. Lett.}\ }\textbf {\bibinfo {volume} {118}},\ \bibinfo {pages}
  {267701} (\bibinfo {year} {2017})}\BibitemShut {NoStop}%
\bibitem [{\citenamefont {Zvyagin}(2022)}]{Zvyagin2022}%
  \BibitemOpen
  \bibfield  {author} {\bibinfo {author} {\bibfnamefont {A.~A.}\ \bibnamefont
  {Zvyagin}},\ }\bibfield  {title} {\enquote {\bibinfo {title} {Charging of
  majorana edge modes caused by interaction: Exact results},}\ }\href {\doibase
  10.1103/PhysRevB.105.115406} {\bibfield  {journal} {\bibinfo  {journal}
  {Phys. Rev. B}\ }\textbf {\bibinfo {volume} {105}},\ \bibinfo {pages}
  {115406} (\bibinfo {year} {2022})}\BibitemShut {NoStop}%
\bibitem [{\citenamefont {Lin}(1990)}]{Lin1990}%
  \BibitemOpen
  \bibfield  {author} {\bibinfo {author} {\bibfnamefont {HQ}~\bibnamefont
  {Lin}},\ }\bibfield  {title} {\enquote {\bibinfo {title} {Exact
  diagonalization of quantum-spin models},}\ }\href@noop {} {\bibfield
  {journal} {\bibinfo  {journal} {Physical Review B}\ }\textbf {\bibinfo
  {volume} {42}},\ \bibinfo {pages} {6561} (\bibinfo {year}
  {1990})}\BibitemShut {NoStop}%
\bibitem [{\citenamefont {Vidal}(2004)}]{Vidal2004}%
  \BibitemOpen
  \bibfield  {author} {\bibinfo {author} {\bibfnamefont {Guifr\'e}\
  \bibnamefont {Vidal}},\ }\bibfield  {title} {\enquote {\bibinfo {title}
  {Efficient simulation of one-dimensional quantum many-body systems},}\ }\href
  {\doibase 10.1103/PhysRevLett.93.040502} {\bibfield  {journal} {\bibinfo
  {journal} {Phys. Rev. Lett.}\ }\textbf {\bibinfo {volume} {93}},\ \bibinfo
  {pages} {040502} (\bibinfo {year} {2004})}\BibitemShut {NoStop}%
\bibitem [{\citenamefont {Schollw\"ock}(2005)}]{Schollwock2005}%
  \BibitemOpen
  \bibfield  {author} {\bibinfo {author} {\bibfnamefont {U.}~\bibnamefont
  {Schollw\"ock}},\ }\bibfield  {title} {\enquote {\bibinfo {title} {The
  density-matrix renormalization group},}\ }\href {\doibase
  10.1103/RevModPhys.77.259} {\bibfield  {journal} {\bibinfo  {journal} {Rev.
  Mod. Phys.}\ }\textbf {\bibinfo {volume} {77}},\ \bibinfo {pages} {259--315}
  (\bibinfo {year} {2005})}\BibitemShut {NoStop}%
\bibitem [{\citenamefont {Kotliar}\ \emph {et~al.}(2006)\citenamefont
  {Kotliar}, \citenamefont {Savrasov}, \citenamefont {Haule}, \citenamefont
  {Oudovenko}, \citenamefont {Parcollet},\ and\ \citenamefont
  {Marianetti}}]{Kotliar2006}%
  \BibitemOpen
  \bibfield  {author} {\bibinfo {author} {\bibfnamefont {G.}~\bibnamefont
  {Kotliar}}, \bibinfo {author} {\bibfnamefont {S.~Y.}\ \bibnamefont
  {Savrasov}}, \bibinfo {author} {\bibfnamefont {K.}~\bibnamefont {Haule}},
  \bibinfo {author} {\bibfnamefont {V.~S.}\ \bibnamefont {Oudovenko}}, \bibinfo
  {author} {\bibfnamefont {O.}~\bibnamefont {Parcollet}}, \ and\ \bibinfo
  {author} {\bibfnamefont {C.~A.}\ \bibnamefont {Marianetti}},\ }\bibfield
  {title} {\enquote {\bibinfo {title} {Electronic structure calculations with
  dynamical mean-field theory},}\ }\href {\doibase 10.1103/RevModPhys.78.865}
  {\bibfield  {journal} {\bibinfo  {journal} {Rev. Mod. Phys.}\ }\textbf
  {\bibinfo {volume} {78}},\ \bibinfo {pages} {865--951} (\bibinfo {year}
  {2006})}\BibitemShut {NoStop}%
\bibitem [{\citenamefont {Stoudenmire}\ \emph {et~al.}(2011)\citenamefont
  {Stoudenmire}, \citenamefont {Alicea}, \citenamefont {Starykh},\ and\
  \citenamefont {Fisher}}]{Stoudenmire2011}%
  \BibitemOpen
  \bibfield  {author} {\bibinfo {author} {\bibfnamefont {E.~M.}\ \bibnamefont
  {Stoudenmire}}, \bibinfo {author} {\bibfnamefont {Jason}\ \bibnamefont
  {Alicea}}, \bibinfo {author} {\bibfnamefont {Oleg~A.}\ \bibnamefont
  {Starykh}}, \ and\ \bibinfo {author} {\bibfnamefont {Matthew~P.A.}\
  \bibnamefont {Fisher}},\ }\bibfield  {title} {\enquote {\bibinfo {title}
  {Interaction effects in topological superconducting wires supporting majorana
  fermions},}\ }\href {\doibase 10.1103/PhysRevB.84.014503} {\bibfield
  {journal} {\bibinfo  {journal} {Phys. Rev. B}\ }\textbf {\bibinfo {volume}
  {84}},\ \bibinfo {pages} {014503} (\bibinfo {year} {2011})}\BibitemShut
  {NoStop}%
\bibitem [{\citenamefont {Silvi}\ \emph {et~al.}(2019)\citenamefont {Silvi},
  \citenamefont {Tschirsich}, \citenamefont {Gerster}, \citenamefont
  {J{\"u}nemann}, \citenamefont {Jaschke}, \citenamefont {Rizzi},\ and\
  \citenamefont {Montangero}}]{Silvi2019}%
  \BibitemOpen
  \bibfield  {author} {\bibinfo {author} {\bibfnamefont {Pietro}\ \bibnamefont
  {Silvi}}, \bibinfo {author} {\bibfnamefont {Ferdinand}\ \bibnamefont
  {Tschirsich}}, \bibinfo {author} {\bibfnamefont {Matthias}\ \bibnamefont
  {Gerster}}, \bibinfo {author} {\bibfnamefont {Johannes}\ \bibnamefont
  {J{\"u}nemann}}, \bibinfo {author} {\bibfnamefont {Daniel}\ \bibnamefont
  {Jaschke}}, \bibinfo {author} {\bibfnamefont {Matteo}\ \bibnamefont {Rizzi}},
  \ and\ \bibinfo {author} {\bibfnamefont {Simone}\ \bibnamefont
  {Montangero}},\ }\bibfield  {title} {\enquote {\bibinfo {title} {The tensor
  networks anthology: Simulation techniques for many-body quantum lattice
  systems},}\ }\href@noop {} {\bibfield  {journal} {\bibinfo  {journal}
  {SciPost Physics Lecture Notes}\ ,\ \bibinfo {pages} {008}} (\bibinfo {year}
  {2019})}\BibitemShut {NoStop}%
\bibitem [{\citenamefont {Tuovinen}(2021)}]{Tuovinen2021}%
  \BibitemOpen
  \bibfield  {author} {\bibinfo {author} {\bibfnamefont {Riku}\ \bibnamefont
  {Tuovinen}},\ }\bibfield  {title} {\enquote {\bibinfo {title} {{Electron
  correlation effects in superconducting nanowires in and out of
  equilibrium}},}\ }\href {\doibase 10.1088/1367-2630/ac1898} {\bibfield
  {journal} {\bibinfo  {journal} {New Journal of Physics}\ }\textbf {\bibinfo
  {volume} {23}} (\bibinfo {year} {2021}),\ 10.1088/1367-2630/ac1898},\ \Eprint
  {http://arxiv.org/abs/2105.06193} {arXiv:2105.06193} \BibitemShut {NoStop}%
\bibitem [{\citenamefont {Esaki}\ \emph {et~al.}(2011)\citenamefont {Esaki},
  \citenamefont {Sato}, \citenamefont {Hasebe},\ and\ \citenamefont
  {Kohmoto}}]{Esaki2011}%
  \BibitemOpen
  \bibfield  {author} {\bibinfo {author} {\bibfnamefont {Kenta}\ \bibnamefont
  {Esaki}}, \bibinfo {author} {\bibfnamefont {Masatoshi}\ \bibnamefont {Sato}},
  \bibinfo {author} {\bibfnamefont {Kazuki}\ \bibnamefont {Hasebe}}, \ and\
  \bibinfo {author} {\bibfnamefont {Mahito}\ \bibnamefont {Kohmoto}},\
  }\bibfield  {title} {\enquote {\bibinfo {title} {Edge states and topological
  phases in non-hermitian systems},}\ }\href {\doibase
  10.1103/PhysRevB.84.205128} {\bibfield  {journal} {\bibinfo  {journal} {Phys.
  Rev. B}\ }\textbf {\bibinfo {volume} {84}},\ \bibinfo {pages} {205128}
  (\bibinfo {year} {2011})}\BibitemShut {NoStop}%
\bibitem [{\citenamefont {Malzard}\ \emph {et~al.}(2015)\citenamefont
  {Malzard}, \citenamefont {Poli},\ and\ \citenamefont
  {Schomerus}}]{Malzard2015}%
  \BibitemOpen
  \bibfield  {author} {\bibinfo {author} {\bibfnamefont {Simon}\ \bibnamefont
  {Malzard}}, \bibinfo {author} {\bibfnamefont {Charles}\ \bibnamefont {Poli}},
  \ and\ \bibinfo {author} {\bibfnamefont {Henning}\ \bibnamefont
  {Schomerus}},\ }\bibfield  {title} {\enquote {\bibinfo {title} {Topologically
  protected defect states in open photonic systems with non-hermitian
  charge-conjugation and parity-time symmetry},}\ }\href {\doibase
  10.1103/PhysRevLett.115.200402} {\bibfield  {journal} {\bibinfo  {journal}
  {Phys. Rev. Lett.}\ }\textbf {\bibinfo {volume} {115}},\ \bibinfo {pages}
  {200402} (\bibinfo {year} {2015})}\BibitemShut {NoStop}%
\bibitem [{\citenamefont {Gong}\ \emph {et~al.}(2018)\citenamefont {Gong},
  \citenamefont {Ashida}, \citenamefont {Kawabata}, \citenamefont {Takasan},
  \citenamefont {Higashikawa},\ and\ \citenamefont {Ueda}}]{Gong2018}%
  \BibitemOpen
  \bibfield  {author} {\bibinfo {author} {\bibfnamefont {Zongping}\
  \bibnamefont {Gong}}, \bibinfo {author} {\bibfnamefont {Yuto}\ \bibnamefont
  {Ashida}}, \bibinfo {author} {\bibfnamefont {Kohei}\ \bibnamefont
  {Kawabata}}, \bibinfo {author} {\bibfnamefont {Kazuaki}\ \bibnamefont
  {Takasan}}, \bibinfo {author} {\bibfnamefont {Sho}\ \bibnamefont
  {Higashikawa}}, \ and\ \bibinfo {author} {\bibfnamefont {Masahito}\
  \bibnamefont {Ueda}},\ }\bibfield  {title} {\enquote {\bibinfo {title}
  {Topological phases of non-hermitian systems},}\ }\href {\doibase
  10.1103/PhysRevX.8.031079} {\bibfield  {journal} {\bibinfo  {journal} {Phys.
  Rev. X}\ }\textbf {\bibinfo {volume} {8}},\ \bibinfo {pages} {031079}
  (\bibinfo {year} {2018})}\BibitemShut {NoStop}%
\bibitem [{\citenamefont {Yao}\ and\ \citenamefont {Wang}(2018)}]{Yao2018}%
  \BibitemOpen
  \bibfield  {author} {\bibinfo {author} {\bibfnamefont {Shunyu}\ \bibnamefont
  {Yao}}\ and\ \bibinfo {author} {\bibfnamefont {Zhong}\ \bibnamefont {Wang}},\
  }\bibfield  {title} {\enquote {\bibinfo {title} {Edge states and topological
  invariants of non-hermitian systems},}\ }\href {\doibase
  10.1103/PhysRevLett.121.086803} {\bibfield  {journal} {\bibinfo  {journal}
  {Phys. Rev. Lett.}\ }\textbf {\bibinfo {volume} {121}},\ \bibinfo {pages}
  {086803} (\bibinfo {year} {2018})}\BibitemShut {NoStop}%
\bibitem [{\citenamefont {Shen}\ \emph {et~al.}(2018)\citenamefont {Shen},
  \citenamefont {Zhen},\ and\ \citenamefont {Fu}}]{Shen2018}%
  \BibitemOpen
  \bibfield  {author} {\bibinfo {author} {\bibfnamefont {Huitao}\ \bibnamefont
  {Shen}}, \bibinfo {author} {\bibfnamefont {Bo}~\bibnamefont {Zhen}}, \ and\
  \bibinfo {author} {\bibfnamefont {Liang}\ \bibnamefont {Fu}},\ }\bibfield
  {title} {\enquote {\bibinfo {title} {Topological band theory for
  non-hermitian hamiltonians},}\ }\href {\doibase
  10.1103/PhysRevLett.120.146402} {\bibfield  {journal} {\bibinfo  {journal}
  {Phys. Rev. Lett.}\ }\textbf {\bibinfo {volume} {120}},\ \bibinfo {pages}
  {146402} (\bibinfo {year} {2018})}\BibitemShut {NoStop}%
\bibitem [{\citenamefont {Kawabata}\ \emph {et~al.}(2019)\citenamefont
  {Kawabata}, \citenamefont {Shiozaki}, \citenamefont {Ueda},\ and\
  \citenamefont {Sato}}]{Kawabata2019}%
  \BibitemOpen
  \bibfield  {author} {\bibinfo {author} {\bibfnamefont {Kohei}\ \bibnamefont
  {Kawabata}}, \bibinfo {author} {\bibfnamefont {Ken}\ \bibnamefont
  {Shiozaki}}, \bibinfo {author} {\bibfnamefont {Masahito}\ \bibnamefont
  {Ueda}}, \ and\ \bibinfo {author} {\bibfnamefont {Masatoshi}\ \bibnamefont
  {Sato}},\ }\bibfield  {title} {\enquote {\bibinfo {title} {Symmetry and
  topology in non-hermitian physics},}\ }\href {\doibase
  10.1103/PhysRevX.9.041015} {\bibfield  {journal} {\bibinfo  {journal} {Phys.
  Rev. X}\ }\textbf {\bibinfo {volume} {9}},\ \bibinfo {pages} {041015}
  (\bibinfo {year} {2019})}\BibitemShut {NoStop}%
\bibitem [{\citenamefont {Yokomizo}\ and\ \citenamefont
  {Murakami}(2019)}]{Yokomizo2019}%
  \BibitemOpen
  \bibfield  {author} {\bibinfo {author} {\bibfnamefont {Kazuki}\ \bibnamefont
  {Yokomizo}}\ and\ \bibinfo {author} {\bibfnamefont {Shuichi}\ \bibnamefont
  {Murakami}},\ }\bibfield  {title} {\enquote {\bibinfo {title} {Non-bloch band
  theory of non-hermitian systems},}\ }\href {\doibase
  10.1103/PhysRevLett.123.066404} {\bibfield  {journal} {\bibinfo  {journal}
  {Phys. Rev. Lett.}\ }\textbf {\bibinfo {volume} {123}},\ \bibinfo {pages}
  {066404} (\bibinfo {year} {2019})}\BibitemShut {NoStop}%
\bibitem [{\citenamefont {Zhou}\ and\ \citenamefont {Lee}(2019)}]{Zhou2019}%
  \BibitemOpen
  \bibfield  {author} {\bibinfo {author} {\bibfnamefont {Hengyun}\ \bibnamefont
  {Zhou}}\ and\ \bibinfo {author} {\bibfnamefont {Jong~Yeon}\ \bibnamefont
  {Lee}},\ }\bibfield  {title} {\enquote {\bibinfo {title} {Periodic table for
  topological bands with non-hermitian symmetries},}\ }\href {\doibase
  10.1103/PhysRevB.99.235112} {\bibfield  {journal} {\bibinfo  {journal} {Phys.
  Rev. B}\ }\textbf {\bibinfo {volume} {99}},\ \bibinfo {pages} {235112}
  (\bibinfo {year} {2019})}\BibitemShut {NoStop}%
\bibitem [{\citenamefont {Borgnia}\ \emph {et~al.}(2020)\citenamefont
  {Borgnia}, \citenamefont {Kruchkov},\ and\ \citenamefont
  {Slager}}]{Borgnia2020}%
  \BibitemOpen
  \bibfield  {author} {\bibinfo {author} {\bibfnamefont {Dan~S.}\ \bibnamefont
  {Borgnia}}, \bibinfo {author} {\bibfnamefont {Alex~Jura}\ \bibnamefont
  {Kruchkov}}, \ and\ \bibinfo {author} {\bibfnamefont {Robert-Jan}\
  \bibnamefont {Slager}},\ }\bibfield  {title} {\enquote {\bibinfo {title}
  {Non-hermitian boundary modes and topology},}\ }\href {\doibase
  10.1103/PhysRevLett.124.056802} {\bibfield  {journal} {\bibinfo  {journal}
  {Phys. Rev. Lett.}\ }\textbf {\bibinfo {volume} {124}},\ \bibinfo {pages}
  {056802} (\bibinfo {year} {2020})}\BibitemShut {NoStop}%
\bibitem [{\citenamefont {Ashida}\ \emph {et~al.}(2020)\citenamefont {Ashida},
  \citenamefont {Gong},\ and\ \citenamefont {Ueda}}]{Ashida2020}%
  \BibitemOpen
  \bibfield  {author} {\bibinfo {author} {\bibfnamefont {Yuto}\ \bibnamefont
  {Ashida}}, \bibinfo {author} {\bibfnamefont {Zongping}\ \bibnamefont {Gong}},
  \ and\ \bibinfo {author} {\bibfnamefont {Masahito}\ \bibnamefont {Ueda}},\
  }\bibfield  {title} {\enquote {\bibinfo {title} {Non-hermitian physics},}\
  }\href {\doibase 10.1080/00018732.2021.1876991} {\bibfield  {journal}
  {\bibinfo  {journal} {Advances in Physics}\ }\textbf {\bibinfo {volume}
  {69}},\ \bibinfo {pages} {249--435} (\bibinfo {year} {2020})}\BibitemShut
  {NoStop}%
\bibitem [{\citenamefont {Sayyad}\ \emph
  {et~al.}(2022{\natexlab{a}})\citenamefont {Sayyad}, \citenamefont
  {Hannukainen},\ and\ \citenamefont {Grushin}}]{Sayyad2022d}%
  \BibitemOpen
  \bibfield  {author} {\bibinfo {author} {\bibfnamefont {Sharareh}\
  \bibnamefont {Sayyad}}, \bibinfo {author} {\bibfnamefont {Julia~D.}\
  \bibnamefont {Hannukainen}}, \ and\ \bibinfo {author} {\bibfnamefont
  {Adolfo~G.}\ \bibnamefont {Grushin}},\ }\bibfield  {title} {\enquote
  {\bibinfo {title} {Non-hermitian chiral anomalies},}\ }\href {\doibase
  10.1103/PhysRevResearch.4.L042004} {\bibfield  {journal} {\bibinfo  {journal}
  {Phys. Rev. Res.}\ }\textbf {\bibinfo {volume} {4}},\ \bibinfo {pages}
  {L042004} (\bibinfo {year} {2022}{\natexlab{a}})}\BibitemShut {NoStop}%
\bibitem [{\citenamefont {Okuma}\ and\ \citenamefont {Sato}(2023)}]{Okuma2023}%
  \BibitemOpen
  \bibfield  {author} {\bibinfo {author} {\bibfnamefont {Nobuyuki}\
  \bibnamefont {Okuma}}\ and\ \bibinfo {author} {\bibfnamefont {Masatoshi}\
  \bibnamefont {Sato}},\ }\bibfield  {title} {\enquote {\bibinfo {title}
  {Non-hermitian topological phenomena: A review},}\ }\href {\doibase
  10.1146/annurev-conmatphys-040521-033133} {\bibfield  {journal} {\bibinfo
  {journal} {Annual Review of Condensed Matter Physics}\ }\textbf {\bibinfo
  {volume} {14}},\ \bibinfo {pages} {null} (\bibinfo {year} {2023})},\ \Eprint
  {http://arxiv.org/abs/https://doi.org/10.1146/annurev-conmatphys-040521-033133}
  {https://doi.org/10.1146/annurev-conmatphys-040521-033133} \BibitemShut
  {NoStop}%
\bibitem [{\citenamefont {Prosen}(2008)}]{Prosen2008}%
  \BibitemOpen
  \bibfield  {author} {\bibinfo {author} {\bibfnamefont {Tomaz}\ \bibnamefont
  {Prosen}},\ }\bibfield  {title} {\enquote {\bibinfo {title} {Third
  quantization: a general method to solve master equations for quadratic open
  fermi systems},}\ }\href {\doibase 10.1088/1367-2630/10/4/043026} {\bibfield
  {journal} {\bibinfo  {journal} {New Journal of Physics}\ }\textbf {\bibinfo
  {volume} {10}},\ \bibinfo {pages} {043026} (\bibinfo {year}
  {2008})}\BibitemShut {NoStop}%
\bibitem [{\citenamefont {Lieu}\ \emph {et~al.}(2020)\citenamefont {Lieu},
  \citenamefont {McGinley},\ and\ \citenamefont {Cooper}}]{Lieu2020}%
  \BibitemOpen
  \bibfield  {author} {\bibinfo {author} {\bibfnamefont {Simon}\ \bibnamefont
  {Lieu}}, \bibinfo {author} {\bibfnamefont {Max}\ \bibnamefont {McGinley}}, \
  and\ \bibinfo {author} {\bibfnamefont {Nigel~R.}\ \bibnamefont {Cooper}},\
  }\bibfield  {title} {\enquote {\bibinfo {title} {Tenfold way for quadratic
  lindbladians},}\ }\href {\doibase 10.1103/PhysRevLett.124.040401} {\bibfield
  {journal} {\bibinfo  {journal} {Phys. Rev. Lett.}\ }\textbf {\bibinfo
  {volume} {124}},\ \bibinfo {pages} {040401} (\bibinfo {year}
  {2020})}\BibitemShut {NoStop}%
\bibitem [{\citenamefont {Sayyad}\ \emph {et~al.}(2021)\citenamefont {Sayyad},
  \citenamefont {Yu}, \citenamefont {Grushin},\ and\ \citenamefont
  {Sieberer}}]{Sayyad2021}%
  \BibitemOpen
  \bibfield  {author} {\bibinfo {author} {\bibfnamefont {Sharareh}\
  \bibnamefont {Sayyad}}, \bibinfo {author} {\bibfnamefont {Jinlong}\
  \bibnamefont {Yu}}, \bibinfo {author} {\bibfnamefont {Adolfo~G.}\
  \bibnamefont {Grushin}}, \ and\ \bibinfo {author} {\bibfnamefont {Lukas~M.}\
  \bibnamefont {Sieberer}},\ }\bibfield  {title} {\enquote {\bibinfo {title}
  {Entanglement spectrum crossings reveal non-hermitian dynamical topology},}\
  }\href {\doibase 10.1103/PhysRevResearch.3.033022} {\bibfield  {journal}
  {\bibinfo  {journal} {Phys. Rev. Res.}\ }\textbf {\bibinfo {volume} {3}},\
  \bibinfo {pages} {033022} (\bibinfo {year} {2021})}\BibitemShut {NoStop}%
\bibitem [{\citenamefont {Jr}\ \emph {et~al.}(2022)\citenamefont {Jr},
  \citenamefont {Miranowicz}, \citenamefont {Chimczak},\ and\ \citenamefont
  {Kowalewska-Kudlaszyk}}]{Perina2022}%
  \BibitemOpen
  \bibfield  {author} {\bibinfo {author} {\bibfnamefont {Jan~Perina}\
  \bibnamefont {Jr}}, \bibinfo {author} {\bibfnamefont {Adam}\ \bibnamefont
  {Miranowicz}}, \bibinfo {author} {\bibfnamefont {Grzegorz}\ \bibnamefont
  {Chimczak}}, \ and\ \bibinfo {author} {\bibfnamefont {Anna}\ \bibnamefont
  {Kowalewska-Kudlaszyk}},\ }\bibfield  {title} {\enquote {\bibinfo {title}
  {Quantum liouvillian exceptional and diabolical points for bosonic fields
  with quadratic hamiltonians: The heisenberg-langevin equation approach},}\
  }\href {\doibase 10.22331/q-2022-12-22-883} {\bibfield  {journal} {\bibinfo
  {journal} {Quantum}\ }\textbf {\bibinfo {volume} {6}},\ \bibinfo {pages}
  {883} (\bibinfo {year} {2022})}\BibitemShut {NoStop}%
\bibitem [{\citenamefont {Yang}\ \emph {et~al.}(2022)\citenamefont {Yang},
  \citenamefont {Jiang},\ and\ \citenamefont {Bergholtz}}]{Yang2022}%
  \BibitemOpen
  \bibfield  {author} {\bibinfo {author} {\bibfnamefont {Fan}\ \bibnamefont
  {Yang}}, \bibinfo {author} {\bibfnamefont {Qing-Dong}\ \bibnamefont {Jiang}},
  \ and\ \bibinfo {author} {\bibfnamefont {Emil~J.}\ \bibnamefont
  {Bergholtz}},\ }\bibfield  {title} {\enquote {\bibinfo {title} {Liouvillian
  skin effect in an exactly solvable model},}\ }\href {\doibase
  10.1103/PhysRevResearch.4.023160} {\bibfield  {journal} {\bibinfo  {journal}
  {Phys. Rev. Res.}\ }\textbf {\bibinfo {volume} {4}},\ \bibinfo {pages}
  {023160} (\bibinfo {year} {2022})}\BibitemShut {NoStop}%
\bibitem [{\citenamefont {Talkington}\ and\ \citenamefont
  {Claassen}(2022)}]{Talkington2022}%
  \BibitemOpen
  \bibfield  {author} {\bibinfo {author} {\bibfnamefont {Spenser}\ \bibnamefont
  {Talkington}}\ and\ \bibinfo {author} {\bibfnamefont {Martin}\ \bibnamefont
  {Claassen}},\ }\bibfield  {title} {\enquote {\bibinfo {title}
  {Dissipation-induced flat bands},}\ }\href {\doibase
  10.1103/PhysRevB.106.L161109} {\bibfield  {journal} {\bibinfo  {journal}
  {Phys. Rev. B}\ }\textbf {\bibinfo {volume} {106}},\ \bibinfo {pages}
  {L161109} (\bibinfo {year} {2022})}\BibitemShut {NoStop}%
\bibitem [{\citenamefont {Starchl}\ and\ \citenamefont
  {Sieberer}(2022)}]{Starchl2022}%
  \BibitemOpen
  \bibfield  {author} {\bibinfo {author} {\bibfnamefont {Elias}\ \bibnamefont
  {Starchl}}\ and\ \bibinfo {author} {\bibfnamefont {Lukas~M.}\ \bibnamefont
  {Sieberer}},\ }\bibfield  {title} {\enquote {\bibinfo {title} {Relaxation to
  a parity-time symmetric generalized gibbs ensemble after a quantum quench in
  a driven-dissipative kitaev chain},}\ }\href {\doibase
  10.1103/PhysRevLett.129.220602} {\bibfield  {journal} {\bibinfo  {journal}
  {Phys. Rev. Lett.}\ }\textbf {\bibinfo {volume} {129}},\ \bibinfo {pages}
  {220602} (\bibinfo {year} {2022})}\BibitemShut {NoStop}%
\bibitem [{\citenamefont {Gneiting}\ \emph {et~al.}(2022)\citenamefont
  {Gneiting}, \citenamefont {Koottandavida}, \citenamefont {Rozhkov},\ and\
  \citenamefont {Nori}}]{Gneiting2022}%
  \BibitemOpen
  \bibfield  {author} {\bibinfo {author} {\bibfnamefont {Clemens}\ \bibnamefont
  {Gneiting}}, \bibinfo {author} {\bibfnamefont {Akshay}\ \bibnamefont
  {Koottandavida}}, \bibinfo {author} {\bibfnamefont {A.~V.}\ \bibnamefont
  {Rozhkov}}, \ and\ \bibinfo {author} {\bibfnamefont {Franco}\ \bibnamefont
  {Nori}},\ }\bibfield  {title} {\enquote {\bibinfo {title} {Unraveling the
  topology of dissipative quantum systems},}\ }\href {\doibase
  10.1103/PhysRevResearch.4.023036} {\bibfield  {journal} {\bibinfo  {journal}
  {Phys. Rev. Res.}\ }\textbf {\bibinfo {volume} {4}},\ \bibinfo {pages}
  {023036} (\bibinfo {year} {2022})}\BibitemShut {NoStop}%
\bibitem [{\citenamefont {Chen}\ \emph {et~al.}(2021)\citenamefont {Chen},
  \citenamefont {Abbasi}, \citenamefont {Joglekar},\ and\ \citenamefont
  {Murch}}]{Chen2021}%
  \BibitemOpen
  \bibfield  {author} {\bibinfo {author} {\bibfnamefont {Weijian}\ \bibnamefont
  {Chen}}, \bibinfo {author} {\bibfnamefont {Maryam}\ \bibnamefont {Abbasi}},
  \bibinfo {author} {\bibfnamefont {Yogesh~N.}\ \bibnamefont {Joglekar}}, \
  and\ \bibinfo {author} {\bibfnamefont {Kater~W.}\ \bibnamefont {Murch}},\
  }\bibfield  {title} {\enquote {\bibinfo {title} {Quantum jumps in the
  non-hermitian dynamics of a superconducting qubit},}\ }\href {\doibase
  10.1103/PhysRevLett.127.140504} {\bibfield  {journal} {\bibinfo  {journal}
  {Phys. Rev. Lett.}\ }\textbf {\bibinfo {volume} {127}},\ \bibinfo {pages}
  {140504} (\bibinfo {year} {2021})}\BibitemShut {NoStop}%
\bibitem [{\citenamefont {Abbasi}\ \emph {et~al.}(2022)\citenamefont {Abbasi},
  \citenamefont {Chen}, \citenamefont {Naghiloo}, \citenamefont {Joglekar},\
  and\ \citenamefont {Murch}}]{Abbasi2022}%
  \BibitemOpen
  \bibfield  {author} {\bibinfo {author} {\bibfnamefont {Maryam}\ \bibnamefont
  {Abbasi}}, \bibinfo {author} {\bibfnamefont {Weijian}\ \bibnamefont {Chen}},
  \bibinfo {author} {\bibfnamefont {Mahdi}\ \bibnamefont {Naghiloo}}, \bibinfo
  {author} {\bibfnamefont {Yogesh~N.}\ \bibnamefont {Joglekar}}, \ and\
  \bibinfo {author} {\bibfnamefont {Kater~W.}\ \bibnamefont {Murch}},\
  }\bibfield  {title} {\enquote {\bibinfo {title} {Topological quantum state
  control through exceptional-point proximity},}\ }\href {\doibase
  10.1103/PhysRevLett.128.160401} {\bibfield  {journal} {\bibinfo  {journal}
  {Phys. Rev. Lett.}\ }\textbf {\bibinfo {volume} {128}},\ \bibinfo {pages}
  {160401} (\bibinfo {year} {2022})}\BibitemShut {NoStop}%
\bibitem [{\citenamefont {Chen}\ \emph {et~al.}(2022)\citenamefont {Chen},
  \citenamefont {Abbasi}, \citenamefont {Ha}, \citenamefont {Erdamar},
  \citenamefont {Joglekar},\ and\ \citenamefont {Murch}}]{Chen2022}%
  \BibitemOpen
  \bibfield  {author} {\bibinfo {author} {\bibfnamefont {Weijian}\ \bibnamefont
  {Chen}}, \bibinfo {author} {\bibfnamefont {Maryam}\ \bibnamefont {Abbasi}},
  \bibinfo {author} {\bibfnamefont {Byung}\ \bibnamefont {Ha}}, \bibinfo
  {author} {\bibfnamefont {Serra}\ \bibnamefont {Erdamar}}, \bibinfo {author}
  {\bibfnamefont {Yogesh~N.}\ \bibnamefont {Joglekar}}, \ and\ \bibinfo
  {author} {\bibfnamefont {Kater~W.}\ \bibnamefont {Murch}},\ }\bibfield
  {title} {\enquote {\bibinfo {title} {Decoherence-induced exceptional points
  in a dissipative superconducting qubit},}\ }\href {\doibase
  10.1103/PhysRevLett.128.110402} {\bibfield  {journal} {\bibinfo  {journal}
  {Phys. Rev. Lett.}\ }\textbf {\bibinfo {volume} {128}},\ \bibinfo {pages}
  {110402} (\bibinfo {year} {2022})}\BibitemShut {NoStop}%
\bibitem [{\citenamefont {Leykam}\ \emph {et~al.}(2017)\citenamefont {Leykam},
  \citenamefont {Bliokh}, \citenamefont {Huang}, \citenamefont {Chong},\ and\
  \citenamefont {Nori}}]{Leykam2017}%
  \BibitemOpen
  \bibfield  {author} {\bibinfo {author} {\bibfnamefont {Daniel}\ \bibnamefont
  {Leykam}}, \bibinfo {author} {\bibfnamefont {Konstantin~Y.}\ \bibnamefont
  {Bliokh}}, \bibinfo {author} {\bibfnamefont {Chunli}\ \bibnamefont {Huang}},
  \bibinfo {author} {\bibfnamefont {Y.~D.}\ \bibnamefont {Chong}}, \ and\
  \bibinfo {author} {\bibfnamefont {Franco}\ \bibnamefont {Nori}},\ }\bibfield
  {title} {\enquote {\bibinfo {title} {Edge modes, degeneracies, and
  topological numbers in non-hermitian systems},}\ }\href {\doibase
  10.1103/PhysRevLett.118.040401} {\bibfield  {journal} {\bibinfo  {journal}
  {Phys. Rev. Lett.}\ }\textbf {\bibinfo {volume} {118}},\ \bibinfo {pages}
  {040401} (\bibinfo {year} {2017})}\BibitemShut {NoStop}%
\bibitem [{\citenamefont {Bergholtz}\ \emph {et~al.}(2021)\citenamefont
  {Bergholtz}, \citenamefont {Budich},\ and\ \citenamefont
  {Kunst}}]{Bergholtz2021}%
  \BibitemOpen
  \bibfield  {author} {\bibinfo {author} {\bibfnamefont {Emil~J.}\ \bibnamefont
  {Bergholtz}}, \bibinfo {author} {\bibfnamefont {Jan~Carl}\ \bibnamefont
  {Budich}}, \ and\ \bibinfo {author} {\bibfnamefont {Flore~K.}\ \bibnamefont
  {Kunst}},\ }\bibfield  {title} {\enquote {\bibinfo {title} {Exceptional
  topology of non-hermitian systems},}\ }\href {\doibase
  10.1103/RevModPhys.93.015005} {\bibfield  {journal} {\bibinfo  {journal}
  {Rev. Mod. Phys.}\ }\textbf {\bibinfo {volume} {93}},\ \bibinfo {pages}
  {015005} (\bibinfo {year} {2021})}\BibitemShut {NoStop}%
\bibitem [{\citenamefont {Sayyad}\ and\ \citenamefont
  {Kunst}(2022)}]{Sayyad2022}%
  \BibitemOpen
  \bibfield  {author} {\bibinfo {author} {\bibfnamefont {Sharareh}\
  \bibnamefont {Sayyad}}\ and\ \bibinfo {author} {\bibfnamefont {Flore~K.}\
  \bibnamefont {Kunst}},\ }\bibfield  {title} {\enquote {\bibinfo {title}
  {Realizing exceptional points of any order in the presence of symmetry},}\
  }\href {\doibase 10.1103/PhysRevResearch.4.023130} {\bibfield  {journal}
  {\bibinfo  {journal} {Phys. Rev. Research}\ }\textbf {\bibinfo {volume}
  {4}},\ \bibinfo {pages} {023130} (\bibinfo {year} {2022})}\BibitemShut
  {NoStop}%
\bibitem [{\citenamefont {Sayyad}\ \emph
  {et~al.}(2022{\natexlab{b}})\citenamefont {Sayyad}, \citenamefont
  {Stalhammar}, \citenamefont {Rodland},\ and\ \citenamefont
  {Kunst}}]{Sayyad2022b}%
  \BibitemOpen
  \bibfield  {author} {\bibinfo {author} {\bibfnamefont {Sharareh}\
  \bibnamefont {Sayyad}}, \bibinfo {author} {\bibfnamefont {Marcus}\
  \bibnamefont {Stalhammar}}, \bibinfo {author} {\bibfnamefont {Lukas}\
  \bibnamefont {Rodland}}, \ and\ \bibinfo {author} {\bibfnamefont {Flore~K}\
  \bibnamefont {Kunst}},\ }\bibfield  {title} {\enquote {\bibinfo {title}
  {Symmetry-protected exceptional and nodal points in non-hermitian systems},}\
  }\href@noop {} {\bibfield  {journal} {\bibinfo  {journal} {arXiv:2204.13945}\
  } (\bibinfo {year} {2022}{\natexlab{b}})}\BibitemShut {NoStop}%
\bibitem [{\citenamefont {Sayyad}(2022)}]{Sayyad2022c}%
  \BibitemOpen
  \bibfield  {author} {\bibinfo {author} {\bibfnamefont {Sharareh}\
  \bibnamefont {Sayyad}},\ }\bibfield  {title} {\enquote {\bibinfo {title}
  {Protection of all nondefective twofold degeneracies by antiunitary
  symmetries in non-hermitian systems},}\ }\href@noop {} {\bibfield  {journal}
  {\bibinfo  {journal} {Physical Review Research}\ }\textbf {\bibinfo {volume}
  {4}},\ \bibinfo {pages} {043213} (\bibinfo {year} {2022})}\BibitemShut
  {NoStop}%
\bibitem [{\citenamefont {Song}\ \emph {et~al.}(2019)\citenamefont {Song},
  \citenamefont {Yao},\ and\ \citenamefont {Wang}}]{Song2019}%
  \BibitemOpen
  \bibfield  {author} {\bibinfo {author} {\bibfnamefont {Fei}\ \bibnamefont
  {Song}}, \bibinfo {author} {\bibfnamefont {Shunyu}\ \bibnamefont {Yao}}, \
  and\ \bibinfo {author} {\bibfnamefont {Zhong}\ \bibnamefont {Wang}},\
  }\bibfield  {title} {\enquote {\bibinfo {title} {Non-hermitian skin effect
  and chiral damping in open quantum systems},}\ }\href {\doibase
  10.1103/PhysRevLett.123.170401} {\bibfield  {journal} {\bibinfo  {journal}
  {Phys. Rev. Lett.}\ }\textbf {\bibinfo {volume} {123}},\ \bibinfo {pages}
  {170401} (\bibinfo {year} {2019})}\BibitemShut {NoStop}%
\bibitem [{\citenamefont {Zhang}\ \emph {et~al.}(2022)\citenamefont {Zhang},
  \citenamefont {Zhang}, \citenamefont {Lu},\ and\ \citenamefont
  {Chen}}]{Zhang2022}%
  \BibitemOpen
  \bibfield  {author} {\bibinfo {author} {\bibfnamefont {Xiujuan}\ \bibnamefont
  {Zhang}}, \bibinfo {author} {\bibfnamefont {Tian}\ \bibnamefont {Zhang}},
  \bibinfo {author} {\bibfnamefont {Ming-Hui}\ \bibnamefont {Lu}}, \ and\
  \bibinfo {author} {\bibfnamefont {Yan-Feng}\ \bibnamefont {Chen}},\
  }\bibfield  {title} {\enquote {\bibinfo {title} {A review on non-hermitian
  skin effect},}\ }\href {\doibase 10.1080/23746149.2022.2109431} {\bibfield
  {journal} {\bibinfo  {journal} {Advances in Physics: X}\ }\textbf {\bibinfo
  {volume} {7}},\ \bibinfo {pages} {2109431} (\bibinfo {year} {2022})},\
  \Eprint {http://arxiv.org/abs/https://doi.org/10.1080/23746149.2022.2109431}
  {https://doi.org/10.1080/23746149.2022.2109431} \BibitemShut {NoStop}%
\bibitem [{\citenamefont {Freund}(1992)}]{Freund1992}%
  \BibitemOpen
  \bibfield  {author} {\bibinfo {author} {\bibfnamefont {Roland~W}\
  \bibnamefont {Freund}},\ }\bibfield  {title} {\enquote {\bibinfo {title}
  {Quasi-kernel polynomials and their use in non-hermitian matrix
  iterations},}\ }\href@noop {} {\bibfield  {journal} {\bibinfo  {journal}
  {Journal of Computational and Applied Mathematics}\ }\textbf {\bibinfo
  {volume} {43}},\ \bibinfo {pages} {135--158} (\bibinfo {year}
  {1992})}\BibitemShut {NoStop}%
\bibitem [{\citenamefont {Freund}\ \emph {et~al.}(1993)\citenamefont {Freund},
  \citenamefont {Gutknecht},\ and\ \citenamefont {Nachtigal}}]{Freund1993}%
  \BibitemOpen
  \bibfield  {author} {\bibinfo {author} {\bibfnamefont {Roland~W}\
  \bibnamefont {Freund}}, \bibinfo {author} {\bibfnamefont {Martin~H}\
  \bibnamefont {Gutknecht}}, \ and\ \bibinfo {author} {\bibfnamefont
  {No{\"e}l~M}\ \bibnamefont {Nachtigal}},\ }\bibfield  {title} {\enquote
  {\bibinfo {title} {An implementation of the look-ahead lanczos algorithm for
  non-hermitian matrices},}\ }\href@noop {} {\bibfield  {journal} {\bibinfo
  {journal} {SIAM journal on scientific computing}\ }\textbf {\bibinfo {volume}
  {14}},\ \bibinfo {pages} {137--158} (\bibinfo {year} {1993})}\BibitemShut
  {NoStop}%
\bibitem [{\citenamefont {Guo}\ \emph {et~al.}(2022)\citenamefont {Guo},
  \citenamefont {Xu}, \citenamefont {Li}, \citenamefont {You},\ and\
  \citenamefont {Yang}}]{Guo2022}%
  \BibitemOpen
  \bibfield  {author} {\bibinfo {author} {\bibfnamefont {Zhen}\ \bibnamefont
  {Guo}}, \bibinfo {author} {\bibfnamefont {Zheng-Tao}\ \bibnamefont {Xu}},
  \bibinfo {author} {\bibfnamefont {Meng}\ \bibnamefont {Li}}, \bibinfo
  {author} {\bibfnamefont {Li}~\bibnamefont {You}}, \ and\ \bibinfo {author}
  {\bibfnamefont {Shuo}\ \bibnamefont {Yang}},\ }\bibfield  {title} {\enquote
  {\bibinfo {title} {Variational matrix product state approach for
  non-hermitian system based on a companion hermitian hamiltonian},}\
  }\href@noop {} {\bibfield  {journal} {\bibinfo  {journal} {arXiv:2210.14858}\
  } (\bibinfo {year} {2022})},\ \Eprint {http://arxiv.org/abs/2210.14858}
  {arXiv:2210.14858} \BibitemShut {NoStop}%
\bibitem [{\citenamefont {Carden}(2011)}]{Carden2011}%
  \BibitemOpen
  \bibfield  {author} {\bibinfo {author} {\bibfnamefont {Russell}\ \bibnamefont
  {Carden}},\ }\href@noop {} {\emph {\bibinfo {title} {Ritz values and Arnoldi
  convergence for non-Hermitian matrices}}}\ (\bibinfo  {publisher} {Rice
  University},\ \bibinfo {year} {2011})\BibitemShut {NoStop}%
\bibitem [{\citenamefont {Zhang}\ and\ \citenamefont {Dai}(2016)}]{Zhang2016z}%
  \BibitemOpen
  \bibfield  {author} {\bibinfo {author} {\bibfnamefont {Jianhua}\ \bibnamefont
  {Zhang}}\ and\ \bibinfo {author} {\bibfnamefont {Hua}\ \bibnamefont {Dai}},\
  }\bibfield  {title} {\enquote {\bibinfo {title} {Global gpbicg method for
  complex non-hermitian linear systems with multiple right-hand sides},}\
  }\href@noop {} {\bibfield  {journal} {\bibinfo  {journal} {Computational and
  Applied Mathematics}\ }\textbf {\bibinfo {volume} {35}},\ \bibinfo {pages}
  {171--185} (\bibinfo {year} {2016})}\BibitemShut {NoStop}%
\bibitem [{\citenamefont {{Chen}}\ \emph {et~al.}(2022)\citenamefont {{Chen}},
  \citenamefont {{Song}},\ and\ \citenamefont {{Lado}}}]{Chen2022a}%
  \BibitemOpen
  \bibfield  {author} {\bibinfo {author} {\bibfnamefont {Guangze}\ \bibnamefont
  {{Chen}}}, \bibinfo {author} {\bibfnamefont {Fei}\ \bibnamefont {{Song}}}, \
  and\ \bibinfo {author} {\bibfnamefont {Jose~L.}\ \bibnamefont {{Lado}}},\
  }\bibfield  {title} {\enquote {\bibinfo {title} {{Topological spin
  excitations in non-Hermitian spin chains with a generalized kernel polynomial
  algorithm}},}\ }\href {\doibase 10.48550/arXiv.2208.06425} {\bibfield
  {journal} {\bibinfo  {journal} {arXiv}\ } (\bibinfo {year} {2022}),\
  10.48550/arXiv.2208.06425},\ \Eprint {http://arxiv.org/abs/2208.06425}
  {arXiv:2208.06425} \BibitemShut {NoStop}%
\bibitem [{\citenamefont {Fukui}\ and\ \citenamefont
  {Kawakami}(1998)}]{Fukui1998}%
  \BibitemOpen
  \bibfield  {author} {\bibinfo {author} {\bibfnamefont {Takahiro}\
  \bibnamefont {Fukui}}\ and\ \bibinfo {author} {\bibfnamefont {Norio}\
  \bibnamefont {Kawakami}},\ }\bibfield  {title} {\enquote {\bibinfo {title}
  {{Breakdown of the Mott insulator: Exact solution of an asymmetric Hubbard
  model}},}\ }\href {\doibase 10.1103/PhysRevB.58.16051} {\bibfield  {journal}
  {\bibinfo  {journal} {Physical Review B - Condensed Matter and Materials
  Physics}\ }\textbf {\bibinfo {volume} {58}},\ \bibinfo {pages} {16051--16056}
  (\bibinfo {year} {1998})},\ \Eprint {http://arxiv.org/abs/9806023}
  {arXiv:9806023 [cond-mat]} \BibitemShut {NoStop}%
\bibitem [{\citenamefont {Bu{\v{c}}a}\ \emph {et~al.}(2020)\citenamefont
  {Bu{\v{c}}a}, \citenamefont {Booker}, \citenamefont {Medenjak},\ and\
  \citenamefont {Jaksch}}]{Buca2020}%
  \BibitemOpen
  \bibfield  {author} {\bibinfo {author} {\bibfnamefont {Berislav}\
  \bibnamefont {Bu{\v{c}}a}}, \bibinfo {author} {\bibfnamefont {Cameron}\
  \bibnamefont {Booker}}, \bibinfo {author} {\bibfnamefont {Marko}\
  \bibnamefont {Medenjak}}, \ and\ \bibinfo {author} {\bibfnamefont {Dieter}\
  \bibnamefont {Jaksch}},\ }\bibfield  {title} {\enquote {\bibinfo {title}
  {Bethe ansatz approach for dissipation: exact solutions of quantum many-body
  dynamics under loss},}\ }\href {\doibase 10.1088/1367-2630/abd124} {\bibfield
   {journal} {\bibinfo  {journal} {New Journal of Physics}\ }\textbf {\bibinfo
  {volume} {22}},\ \bibinfo {pages} {123040} (\bibinfo {year}
  {2020})}\BibitemShut {NoStop}%
\bibitem [{\citenamefont {Zhang}\ and\ \citenamefont {Song}(2021)}]{Zhang2021}%
  \BibitemOpen
  \bibfield  {author} {\bibinfo {author} {\bibfnamefont {X.~Z.}\ \bibnamefont
  {Zhang}}\ and\ \bibinfo {author} {\bibfnamefont {Z.}~\bibnamefont {Song}},\
  }\bibfield  {title} {\enquote {\bibinfo {title} {$\ensuremath{\eta}$-pairing
  ground states in the non-hermitian hubbard model},}\ }\href {\doibase
  10.1103/PhysRevB.103.235153} {\bibfield  {journal} {\bibinfo  {journal}
  {Phys. Rev. B}\ }\textbf {\bibinfo {volume} {103}},\ \bibinfo {pages}
  {235153} (\bibinfo {year} {2021})}\BibitemShut {NoStop}%
\bibitem [{\citenamefont {Nakagawa}\ \emph {et~al.}(2021)\citenamefont
  {Nakagawa}, \citenamefont {Kawakami},\ and\ \citenamefont
  {Ueda}}]{Nakagawa2021}%
  \BibitemOpen
  \bibfield  {author} {\bibinfo {author} {\bibfnamefont {Masaya}\ \bibnamefont
  {Nakagawa}}, \bibinfo {author} {\bibfnamefont {Norio}\ \bibnamefont
  {Kawakami}}, \ and\ \bibinfo {author} {\bibfnamefont {Masahito}\ \bibnamefont
  {Ueda}},\ }\bibfield  {title} {\enquote {\bibinfo {title} {Exact liouvillian
  spectrum of a one-dimensional dissipative hubbard model},}\ }\href {\doibase
  10.1103/PhysRevLett.126.110404} {\bibfield  {journal} {\bibinfo  {journal}
  {Phys. Rev. Lett.}\ }\textbf {\bibinfo {volume} {126}},\ \bibinfo {pages}
  {110404} (\bibinfo {year} {2021})}\BibitemShut {NoStop}%
\bibitem [{\citenamefont {Yoshida}\ and\ \citenamefont
  {Katsura}(2022)}]{Yoshida2022}%
  \BibitemOpen
  \bibfield  {author} {\bibinfo {author} {\bibfnamefont {Hironobu}\
  \bibnamefont {Yoshida}}\ and\ \bibinfo {author} {\bibfnamefont {Hosho}\
  \bibnamefont {Katsura}},\ }\bibfield  {title} {\enquote {\bibinfo {title}
  {{Exact analysis of the Liouvillian gap and dynamics in the dissipative
  SU($N$) Fermi-Hubbard model}},}\ }\href {http://arxiv.org/abs/2209.03743} {\
  ,\ \bibinfo {pages} {1--7} (\bibinfo {year} {2022})},\ \Eprint
  {http://arxiv.org/abs/2209.03743} {arXiv:2209.03743} \BibitemShut {NoStop}%
\bibitem [{\citenamefont {Hyart}\ and\ \citenamefont {Lado}(2022)}]{Hyart2022}%
  \BibitemOpen
  \bibfield  {author} {\bibinfo {author} {\bibfnamefont {Timo}\ \bibnamefont
  {Hyart}}\ and\ \bibinfo {author} {\bibfnamefont {J.~L.}\ \bibnamefont
  {Lado}},\ }\bibfield  {title} {\enquote {\bibinfo {title} {Non-hermitian
  many-body topological excitations in interacting quantum dots},}\ }\href
  {\doibase 10.1103/PhysRevResearch.4.L012006} {\bibfield  {journal} {\bibinfo
  {journal} {Phys. Rev. Res.}\ }\textbf {\bibinfo {volume} {4}},\ \bibinfo
  {pages} {L012006} (\bibinfo {year} {2022})}\BibitemShut {NoStop}%
\bibitem [{\citenamefont {Yamamoto}\ \emph {et~al.}(2022)\citenamefont
  {Yamamoto}, \citenamefont {Nakagawa}, \citenamefont {Tezuka}, \citenamefont
  {Ueda},\ and\ \citenamefont {Kawakami}}]{Yamamoto2022}%
  \BibitemOpen
  \bibfield  {author} {\bibinfo {author} {\bibfnamefont {Kazuki}\ \bibnamefont
  {Yamamoto}}, \bibinfo {author} {\bibfnamefont {Masaya}\ \bibnamefont
  {Nakagawa}}, \bibinfo {author} {\bibfnamefont {Masaki}\ \bibnamefont
  {Tezuka}}, \bibinfo {author} {\bibfnamefont {Masahito}\ \bibnamefont {Ueda}},
  \ and\ \bibinfo {author} {\bibfnamefont {Norio}\ \bibnamefont {Kawakami}},\
  }\bibfield  {title} {\enquote {\bibinfo {title} {Universal properties of
  dissipative tomonaga-luttinger liquids: Case study of a non-hermitian xxz
  spin chain},}\ }\href {\doibase 10.1103/PhysRevB.105.205125} {\bibfield
  {journal} {\bibinfo  {journal} {Phys. Rev. B}\ }\textbf {\bibinfo {volume}
  {105}},\ \bibinfo {pages} {205125} (\bibinfo {year} {2022})}\BibitemShut
  {NoStop}%
\bibitem [{\citenamefont {Wang}\ \emph {et~al.}(2022)\citenamefont {Wang},
  \citenamefont {You},\ and\ \citenamefont {Sun}}]{Wang2022}%
  \BibitemOpen
  \bibfield  {author} {\bibinfo {author} {\bibfnamefont {Ya-Nan}\ \bibnamefont
  {Wang}}, \bibinfo {author} {\bibfnamefont {Wen-Long}\ \bibnamefont {You}}, \
  and\ \bibinfo {author} {\bibfnamefont {Gaoyong}\ \bibnamefont {Sun}},\
  }\bibfield  {title} {\enquote {\bibinfo {title} {Quantum criticality in
  interacting bosonic kitaev-hubbard models},}\ }\href {\doibase
  10.1103/PhysRevA.106.053315} {\bibfield  {journal} {\bibinfo  {journal}
  {Phys. Rev. A}\ }\textbf {\bibinfo {volume} {106}},\ \bibinfo {pages}
  {053315} (\bibinfo {year} {2022})}\BibitemShut {NoStop}%
\bibitem [{\citenamefont {Lewenstein}\ \emph {et~al.}(2007)\citenamefont
  {Lewenstein}, \citenamefont {Sanpera}, \citenamefont {Ahufinger},
  \citenamefont {Damski}, \citenamefont {Sen(De)},\ and\ \citenamefont
  {Sen}}]{Lewenstein2007}%
  \BibitemOpen
  \bibfield  {author} {\bibinfo {author} {\bibfnamefont {Maciej}\ \bibnamefont
  {Lewenstein}}, \bibinfo {author} {\bibfnamefont {Anna}\ \bibnamefont
  {Sanpera}}, \bibinfo {author} {\bibfnamefont {Veronica}\ \bibnamefont
  {Ahufinger}}, \bibinfo {author} {\bibfnamefont {Bogdan}\ \bibnamefont
  {Damski}}, \bibinfo {author} {\bibfnamefont {Aditi}\ \bibnamefont {Sen(De)}},
  \ and\ \bibinfo {author} {\bibfnamefont {Ujjwal}\ \bibnamefont {Sen}},\
  }\bibfield  {title} {\enquote {\bibinfo {title} {Ultracold atomic gases in
  optical lattices: mimicking condensed matter physics and beyond},}\ }\href
  {\doibase 10.1080/00018730701223200} {\bibfield  {journal} {\bibinfo
  {journal} {Advances in Physics}\ }\textbf {\bibinfo {volume} {56}},\ \bibinfo
  {pages} {243--379} (\bibinfo {year} {2007})},\ \Eprint
  {http://arxiv.org/abs/https://doi.org/10.1080/00018730701223200}
  {https://doi.org/10.1080/00018730701223200} \BibitemShut {NoStop}%
\bibitem [{\citenamefont {{Zhang}}\ \emph {et~al.}(2016)\citenamefont
  {{Zhang}}, \citenamefont {{Lou}}, \citenamefont {{Li}}, \citenamefont
  {{Reno}}, \citenamefont {{Pan}}, \citenamefont {{Watson}}, \citenamefont
  {{Manfra}},\ and\ \citenamefont {{Kono}}}]{Zhang2016}%
  \BibitemOpen
  \bibfield  {author} {\bibinfo {author} {\bibfnamefont {Qi}~\bibnamefont
  {{Zhang}}}, \bibinfo {author} {\bibfnamefont {Minhan}\ \bibnamefont {{Lou}}},
  \bibinfo {author} {\bibfnamefont {Xinwei}\ \bibnamefont {{Li}}}, \bibinfo
  {author} {\bibfnamefont {John~L.}\ \bibnamefont {{Reno}}}, \bibinfo {author}
  {\bibfnamefont {Wei}\ \bibnamefont {{Pan}}}, \bibinfo {author} {\bibfnamefont
  {John~D.}\ \bibnamefont {{Watson}}}, \bibinfo {author} {\bibfnamefont
  {Michael~J.}\ \bibnamefont {{Manfra}}}, \ and\ \bibinfo {author}
  {\bibfnamefont {Junichiro}\ \bibnamefont {{Kono}}},\ }\bibfield  {title}
  {\enquote {\bibinfo {title} {{Collective non-perturbative coupling of 2D
  electrons with high-quality-factor terahertz cavity photons}},}\ }\href
  {\doibase 10.1038/nphys3850} {\bibfield  {journal} {\bibinfo  {journal}
  {Nature Physics}\ }\textbf {\bibinfo {volume} {12}},\ \bibinfo {pages}
  {1005--1011} (\bibinfo {year} {2016})},\ \Eprint
  {http://arxiv.org/abs/1604.08297} {arXiv:1604.08297 [cond-mat.quant-gas]}
  \BibitemShut {NoStop}%
\bibitem [{\citenamefont {Gross}\ and\ \citenamefont
  {Bloch}(2017)}]{Gross2017}%
  \BibitemOpen
  \bibfield  {author} {\bibinfo {author} {\bibfnamefont {Christian}\
  \bibnamefont {Gross}}\ and\ \bibinfo {author} {\bibfnamefont {Immanuel}\
  \bibnamefont {Bloch}},\ }\bibfield  {title} {\enquote {\bibinfo {title}
  {Quantum simulations with ultracold atoms in optical lattices},}\ }\href
  {\doibase 10.1126/science.aal3837} {\bibfield  {journal} {\bibinfo  {journal}
  {Science}\ }\textbf {\bibinfo {volume} {357}},\ \bibinfo {pages} {995--1001}
  (\bibinfo {year} {2017})},\ \Eprint
  {http://arxiv.org/abs/https://www.science.org/doi/pdf/10.1126/science.aal3837}
  {https://www.science.org/doi/pdf/10.1126/science.aal3837} \BibitemShut
  {NoStop}%
\bibitem [{\citenamefont {Rosso}\ \emph {et~al.}(2022)\citenamefont {Rosso},
  \citenamefont {Mazza},\ and\ \citenamefont {Biella}}]{Rosso2022}%
  \BibitemOpen
  \bibfield  {author} {\bibinfo {author} {\bibfnamefont {Lorenzo}\ \bibnamefont
  {Rosso}}, \bibinfo {author} {\bibfnamefont {Leonardo}\ \bibnamefont {Mazza}},
  \ and\ \bibinfo {author} {\bibfnamefont {Alberto}\ \bibnamefont {Biella}},\
  }\bibfield  {title} {\enquote {\bibinfo {title} {{Eightfold way to dark
  states in SU(3) cold gases with two-body losses}},}\ }\href {\doibase
  10.1103/PhysRevA.105.L051302} {\bibfield  {journal} {\bibinfo  {journal}
  {Physical Review A}\ }\textbf {\bibinfo {volume} {105}},\ \bibinfo {pages}
  {1--6} (\bibinfo {year} {2022})}\BibitemShut {NoStop}%
\bibitem [{\citenamefont {Rainis}\ and\ \citenamefont
  {Loss}(2012)}]{Rainis2012}%
  \BibitemOpen
  \bibfield  {author} {\bibinfo {author} {\bibfnamefont {Diego}\ \bibnamefont
  {Rainis}}\ and\ \bibinfo {author} {\bibfnamefont {Daniel}\ \bibnamefont
  {Loss}},\ }\bibfield  {title} {\enquote {\bibinfo {title} {Majorana qubit
  decoherence by quasiparticle poisoning},}\ }\href {\doibase
  10.1103/PhysRevB.85.174533} {\bibfield  {journal} {\bibinfo  {journal} {Phys.
  Rev. B}\ }\textbf {\bibinfo {volume} {85}},\ \bibinfo {pages} {174533}
  (\bibinfo {year} {2012})}\BibitemShut {NoStop}%
\bibitem [{\citenamefont {Lieu}(2019)}]{Lieu2019}%
  \BibitemOpen
  \bibfield  {author} {\bibinfo {author} {\bibfnamefont {Simon}\ \bibnamefont
  {Lieu}},\ }\bibfield  {title} {\enquote {\bibinfo {title} {Non-hermitian
  majorana modes protect degenerate steady states},}\ }\href {\doibase
  10.1103/PhysRevB.100.085110} {\bibfield  {journal} {\bibinfo  {journal}
  {Phys. Rev. B}\ }\textbf {\bibinfo {volume} {100}},\ \bibinfo {pages}
  {085110} (\bibinfo {year} {2019})}\BibitemShut {NoStop}%
\bibitem [{\citenamefont {Sakaguchi}\ \emph {et~al.}(2022)\citenamefont
  {Sakaguchi}, \citenamefont {Nishijima},\ and\ \citenamefont
  {Takane}}]{Sakaguchi2022}%
  \BibitemOpen
  \bibfield  {author} {\bibinfo {author} {\bibfnamefont {Tetsuro}\ \bibnamefont
  {Sakaguchi}}, \bibinfo {author} {\bibfnamefont {Hiroto}\ \bibnamefont
  {Nishijima}}, \ and\ \bibinfo {author} {\bibfnamefont {Yositake}\
  \bibnamefont {Takane}},\ }\bibfield  {title} {\enquote {\bibinfo {title}
  {Bulk–boundary correspondence and boundary zero modes in a non-hermitian
  kitaev chain model},}\ }\href {\doibase 10.7566/JPSJ.91.124711} {\bibfield
  {journal} {\bibinfo  {journal} {Journal of the Physical Society of Japan}\
  }\textbf {\bibinfo {volume} {91}},\ \bibinfo {pages} {124711} (\bibinfo
  {year} {2022})},\ \Eprint
  {http://arxiv.org/abs/https://doi.org/10.7566/JPSJ.91.124711}
  {https://doi.org/10.7566/JPSJ.91.124711} \BibitemShut {NoStop}%
\bibitem [{\citenamefont {Hassler}\ and\ \citenamefont
  {Schuricht}(2012)}]{Hassler2012}%
  \BibitemOpen
  \bibfield  {author} {\bibinfo {author} {\bibfnamefont {Fabian}\ \bibnamefont
  {Hassler}}\ and\ \bibinfo {author} {\bibfnamefont {Dirk}\ \bibnamefont
  {Schuricht}},\ }\bibfield  {title} {\enquote {\bibinfo {title} {Strongly
  interacting majorana modes in an array of josephson junctions},}\ }\href
  {\doibase 10.1088/1367-2630/14/12/125018} {\bibfield  {journal} {\bibinfo
  {journal} {New Journal of Physics}\ }\textbf {\bibinfo {volume} {14}},\
  \bibinfo {pages} {125018} (\bibinfo {year} {2012})}\BibitemShut {NoStop}%
\bibitem [{\citenamefont {Joglekar}\ and\ \citenamefont
  {Harter}(2018)}]{Joglekar2018}%
  \BibitemOpen
  \bibfield  {author} {\bibinfo {author} {\bibfnamefont {Yogesh~N.}\
  \bibnamefont {Joglekar}}\ and\ \bibinfo {author} {\bibfnamefont {Andrew~K.}\
  \bibnamefont {Harter}},\ }\bibfield  {title} {\enquote {\bibinfo {title}
  {Passive parity-time-symmetry-breaking transitions without exceptional points
  in dissipative photonic systems},}\ }\href {\doibase 10.1364/PRJ.6.000A51}
  {\bibfield  {journal} {\bibinfo  {journal} {Photon. Res.}\ }\textbf {\bibinfo
  {volume} {6}},\ \bibinfo {pages} {A51--A57} (\bibinfo {year}
  {2018})}\BibitemShut {NoStop}%
\bibitem [{SuppMat()}]{SuppMat}%
  \BibitemOpen
  \bibinfo {note} {The Supplemental Material includes details on mapping the
  non-Hermitian interacting Kitaev chain into a noninteracting Hamiltonian and
  discussions on the thermodynamic limits of topological modes and topological
  boundaries.}\BibitemShut {Stop}%
\bibitem [{\citenamefont {Kitaev}(2001)}]{Kitaev2001}%
  \BibitemOpen
  \bibfield  {author} {\bibinfo {author} {\bibfnamefont {A~Yu}\ \bibnamefont
  {Kitaev}},\ }\bibfield  {title} {\enquote {\bibinfo {title} {Unpaired
  majorana fermions in quantum wires},}\ }\href {\doibase
  10.1070/1063-7869/44/10s/s29} {\bibfield  {journal} {\bibinfo  {journal}
  {Physics-Uspekhi}\ }\textbf {\bibinfo {volume} {44}},\ \bibinfo {pages}
  {131--136} (\bibinfo {year} {2001})}\BibitemShut {NoStop}%
\bibitem [{\citenamefont {Goldstein}\ and\ \citenamefont
  {Chamon}(2012)}]{Goldstein2012}%
  \BibitemOpen
  \bibfield  {author} {\bibinfo {author} {\bibfnamefont {G.}~\bibnamefont
  {Goldstein}}\ and\ \bibinfo {author} {\bibfnamefont {C.}~\bibnamefont
  {Chamon}},\ }\bibfield  {title} {\enquote {\bibinfo {title} {{Exact zero
  modes in closed systems of interacting fermions}},}\ }\href {\doibase
  10.1103/PhysRevB.86.115122} {\bibfield  {journal} {\bibinfo  {journal}
  {Physical Review B - Condensed Matter and Materials Physics}\ }\textbf
  {\bibinfo {volume} {86}},\ \bibinfo {pages} {1--5} (\bibinfo {year}
  {2012})},\ \Eprint {http://arxiv.org/abs/1108.1734} {arXiv:1108.1734}
  \BibitemShut {NoStop}%
\bibitem [{\citenamefont {Yang}\ and\ \citenamefont
  {Feldman}(2014)}]{Yang2014}%
  \BibitemOpen
  \bibfield  {author} {\bibinfo {author} {\bibfnamefont {Guang}\ \bibnamefont
  {Yang}}\ and\ \bibinfo {author} {\bibfnamefont {D.~E.}\ \bibnamefont
  {Feldman}},\ }\bibfield  {title} {\enquote {\bibinfo {title} {{Exact zero
  modes and decoherence in systems of interacting Majorana fermions}},}\ }\href
  {\doibase 10.1103/PhysRevB.89.035136} {\bibfield  {journal} {\bibinfo
  {journal} {Physical Review B - Condensed Matter and Materials Physics}\
  }\textbf {\bibinfo {volume} {89}},\ \bibinfo {pages} {1--12} (\bibinfo {year}
  {2014})},\ \Eprint {http://arxiv.org/abs/1311.7604} {arXiv:1311.7604}
  \BibitemShut {NoStop}%
\bibitem [{\citenamefont {Kells}(2015)}]{Kells2015}%
  \BibitemOpen
  \bibfield  {author} {\bibinfo {author} {\bibfnamefont {G.}~\bibnamefont
  {Kells}},\ }\bibfield  {title} {\enquote {\bibinfo {title} {{Multiparticle
  content of Majorana zero modes in the interacting p -wave wire}},}\ }\href
  {\doibase 10.1103/PhysRevB.92.155434} {\bibfield  {journal} {\bibinfo
  {journal} {Physical Review B - Condensed Matter and Materials Physics}\
  }\textbf {\bibinfo {volume} {92}},\ \bibinfo {pages} {1--15} (\bibinfo {year}
  {2015})},\ \Eprint {http://arxiv.org/abs/1507.06539} {arXiv:1507.06539}
  \BibitemShut {NoStop}%
\bibitem [{\citenamefont {McGinley}\ \emph {et~al.}(2017)\citenamefont
  {McGinley}, \citenamefont {Knolle},\ and\ \citenamefont
  {Nunnenkamp}}]{McGinley2017}%
  \BibitemOpen
  \bibfield  {author} {\bibinfo {author} {\bibfnamefont {Max}\ \bibnamefont
  {McGinley}}, \bibinfo {author} {\bibfnamefont {Johannes}\ \bibnamefont
  {Knolle}}, \ and\ \bibinfo {author} {\bibfnamefont {Andreas}\ \bibnamefont
  {Nunnenkamp}},\ }\bibfield  {title} {\enquote {\bibinfo {title} {{Robustness
  of Majorana edge modes and topological order: Exact results for the symmetric
  interacting Kitaev chain with disorder}},}\ }\href {\doibase
  10.1103/PhysRevB.96.241113} {\bibfield  {journal} {\bibinfo  {journal}
  {Physical Review B}\ }\textbf {\bibinfo {volume} {96}},\ \bibinfo {pages}
  {1--5} (\bibinfo {year} {2017})},\ \Eprint {http://arxiv.org/abs/1706.10249}
  {arXiv:1706.10249} \BibitemShut {NoStop}%
\bibitem [{\citenamefont {Fendley}(2012)}]{Fendley2012}%
  \BibitemOpen
  \bibfield  {author} {\bibinfo {author} {\bibfnamefont {Paul}\ \bibnamefont
  {Fendley}},\ }\bibfield  {title} {\enquote {\bibinfo {title} {{Parafermionic
  edge zero modes in Zn-invariant spin chains}},}\ }\href {\doibase
  10.1088/1742-5468/2012/11/P11020} {\bibfield  {journal} {\bibinfo  {journal}
  {Journal of Statistical Mechanics: Theory and Experiment}\ }\textbf {\bibinfo
  {volume} {2012}} (\bibinfo {year} {2012}),\
  10.1088/1742-5468/2012/11/P11020},\ \Eprint {http://arxiv.org/abs/1209.0472}
  {arXiv:1209.0472} \BibitemShut {NoStop}%
\bibitem [{\citenamefont {Mahyaeh}\ and\ \citenamefont
  {Ardonne}(2020)}]{Mahyaeh2020}%
  \BibitemOpen
  \bibfield  {author} {\bibinfo {author} {\bibfnamefont {Iman}\ \bibnamefont
  {Mahyaeh}}\ and\ \bibinfo {author} {\bibfnamefont {Eddy}\ \bibnamefont
  {Ardonne}},\ }\bibfield  {title} {\enquote {\bibinfo {title} {{Study of the
  phase diagram of the Kitaev-Hubbard chain}},}\ }\href {\doibase
  10.1103/PhysRevB.101.085125} {\bibfield  {journal} {\bibinfo  {journal}
  {Physical Review B}\ }\textbf {\bibinfo {volume} {101}},\ \bibinfo {pages}
  {85125} (\bibinfo {year} {2020})}\BibitemShut {NoStop}%
\bibitem [{\citenamefont {Drost}\ \emph {et~al.}(2017)\citenamefont {Drost},
  \citenamefont {Ojanen}, \citenamefont {Harju},\ and\ \citenamefont
  {Liljeroth}}]{Drost2017}%
  \BibitemOpen
  \bibfield  {author} {\bibinfo {author} {\bibfnamefont {Robert}\ \bibnamefont
  {Drost}}, \bibinfo {author} {\bibfnamefont {Teemu}\ \bibnamefont {Ojanen}},
  \bibinfo {author} {\bibfnamefont {Ari}\ \bibnamefont {Harju}}, \ and\
  \bibinfo {author} {\bibfnamefont {Peter}\ \bibnamefont {Liljeroth}},\
  }\bibfield  {title} {\enquote {\bibinfo {title} {Topological states in
  engineered atomic lattices},}\ }\href {\doibase 10.1038/nphys4080} {\bibfield
   {journal} {\bibinfo  {journal} {Nature Physics}\ }\textbf {\bibinfo {volume}
  {13}},\ \bibinfo {pages} {668--671} (\bibinfo {year} {2017})}\BibitemShut
  {NoStop}%
\bibitem [{\citenamefont {Kempkes}\ \emph {et~al.}(2019)\citenamefont
  {Kempkes}, \citenamefont {Slot}, \citenamefont {van~den Broeke},
  \citenamefont {Capiod}, \citenamefont {Benalcazar}, \citenamefont
  {Vanmaekelbergh}, \citenamefont {Bercioux}, \citenamefont {Swart},\ and\
  \citenamefont {Smith}}]{Kempkes2019}%
  \BibitemOpen
  \bibfield  {author} {\bibinfo {author} {\bibfnamefont {S.~N.}\ \bibnamefont
  {Kempkes}}, \bibinfo {author} {\bibfnamefont {M.~R.}\ \bibnamefont {Slot}},
  \bibinfo {author} {\bibfnamefont {J.~J.}\ \bibnamefont {van~den Broeke}},
  \bibinfo {author} {\bibfnamefont {P.}~\bibnamefont {Capiod}}, \bibinfo
  {author} {\bibfnamefont {W.~A.}\ \bibnamefont {Benalcazar}}, \bibinfo
  {author} {\bibfnamefont {D.}~\bibnamefont {Vanmaekelbergh}}, \bibinfo
  {author} {\bibfnamefont {D.}~\bibnamefont {Bercioux}}, \bibinfo {author}
  {\bibfnamefont {I.}~\bibnamefont {Swart}}, \ and\ \bibinfo {author}
  {\bibfnamefont {C.~Morais}\ \bibnamefont {Smith}},\ }\bibfield  {title}
  {\enquote {\bibinfo {title} {Robust zero-energy modes in an electronic
  higher-order topological insulator},}\ }\href {\doibase
  10.1038/s41563-019-0483-4} {\bibfield  {journal} {\bibinfo  {journal} {Nature
  Materials}\ }\textbf {\bibinfo {volume} {18}},\ \bibinfo {pages} {1292--1297}
  (\bibinfo {year} {2019})}\BibitemShut {NoStop}%
\bibitem [{\citenamefont {Kempkes}\ \emph {et~al.}(2018)\citenamefont
  {Kempkes}, \citenamefont {Slot}, \citenamefont {Freeney}, \citenamefont
  {Zevenhuizen}, \citenamefont {Vanmaekelbergh}, \citenamefont {Swart},\ and\
  \citenamefont {Smith}}]{Kempkes2018}%
  \BibitemOpen
  \bibfield  {author} {\bibinfo {author} {\bibfnamefont {S.~N.}\ \bibnamefont
  {Kempkes}}, \bibinfo {author} {\bibfnamefont {M.~R.}\ \bibnamefont {Slot}},
  \bibinfo {author} {\bibfnamefont {S.~E.}\ \bibnamefont {Freeney}}, \bibinfo
  {author} {\bibfnamefont {S.~J.~M.}\ \bibnamefont {Zevenhuizen}}, \bibinfo
  {author} {\bibfnamefont {D.}~\bibnamefont {Vanmaekelbergh}}, \bibinfo
  {author} {\bibfnamefont {I.}~\bibnamefont {Swart}}, \ and\ \bibinfo {author}
  {\bibfnamefont {C.~Morais}\ \bibnamefont {Smith}},\ }\bibfield  {title}
  {\enquote {\bibinfo {title} {Design and characterization of electrons in a
  fractal geometry},}\ }\href {\doibase 10.1038/s41567-018-0328-0} {\bibfield
  {journal} {\bibinfo  {journal} {Nature Physics}\ }\textbf {\bibinfo {volume}
  {15}},\ \bibinfo {pages} {127--131} (\bibinfo {year} {2018})}\BibitemShut
  {NoStop}%
\bibitem [{\citenamefont {Huda}\ \emph {et~al.}(2020)\citenamefont {Huda},
  \citenamefont {Kezilebieke}, \citenamefont {Ojanen}, \citenamefont {Drost},\
  and\ \citenamefont {Liljeroth}}]{Huda2020}%
  \BibitemOpen
  \bibfield  {author} {\bibinfo {author} {\bibfnamefont {Md~Nurul}\
  \bibnamefont {Huda}}, \bibinfo {author} {\bibfnamefont {Shawulienu}\
  \bibnamefont {Kezilebieke}}, \bibinfo {author} {\bibfnamefont {Teemu}\
  \bibnamefont {Ojanen}}, \bibinfo {author} {\bibfnamefont {Robert}\
  \bibnamefont {Drost}}, \ and\ \bibinfo {author} {\bibfnamefont {Peter}\
  \bibnamefont {Liljeroth}},\ }\bibfield  {title} {\enquote {\bibinfo {title}
  {Tuneable topological domain wall states in engineered atomic chains},}\
  }\href {\doibase 10.1038/s41535-020-0219-3} {\bibfield  {journal} {\bibinfo
  {journal} {npj Quantum Materials}\ }\textbf {\bibinfo {volume} {5}} (\bibinfo
  {year} {2020}),\ 10.1038/s41535-020-0219-3}\BibitemShut {NoStop}%
\bibitem [{\citenamefont {Dvir}\ \emph {et~al.}(2023)\citenamefont {Dvir},
  \citenamefont {Wang}, \citenamefont {van Loo}, \citenamefont {Liu},
  \citenamefont {Mazur}, \citenamefont {Bordin}, \citenamefont {ten Haaf},
  \citenamefont {Wang}, \citenamefont {van Driel}, \citenamefont {Zatelli},
  \citenamefont {Li}, \citenamefont {Malinowski}, \citenamefont {Gazibegovic},
  \citenamefont {Badawy}, \citenamefont {Bakkers}, \citenamefont {Wimmer},\
  and\ \citenamefont {Kouwenhoven}}]{Dvir2023}%
  \BibitemOpen
  \bibfield  {author} {\bibinfo {author} {\bibfnamefont {Tom}\ \bibnamefont
  {Dvir}}, \bibinfo {author} {\bibfnamefont {Guanzhong}\ \bibnamefont {Wang}},
  \bibinfo {author} {\bibfnamefont {Nick}\ \bibnamefont {van Loo}}, \bibinfo
  {author} {\bibfnamefont {Chun-Xiao}\ \bibnamefont {Liu}}, \bibinfo {author}
  {\bibfnamefont {Grzegorz~P.}\ \bibnamefont {Mazur}}, \bibinfo {author}
  {\bibfnamefont {Alberto}\ \bibnamefont {Bordin}}, \bibinfo {author}
  {\bibfnamefont {Sebastiaan L.~D.}\ \bibnamefont {ten Haaf}}, \bibinfo
  {author} {\bibfnamefont {Ji-Yin}\ \bibnamefont {Wang}}, \bibinfo {author}
  {\bibfnamefont {David}\ \bibnamefont {van Driel}}, \bibinfo {author}
  {\bibfnamefont {Francesco}\ \bibnamefont {Zatelli}}, \bibinfo {author}
  {\bibfnamefont {Xiang}\ \bibnamefont {Li}}, \bibinfo {author} {\bibfnamefont
  {Filip~K.}\ \bibnamefont {Malinowski}}, \bibinfo {author} {\bibfnamefont
  {Sasa}\ \bibnamefont {Gazibegovic}}, \bibinfo {author} {\bibfnamefont
  {Ghada}\ \bibnamefont {Badawy}}, \bibinfo {author} {\bibfnamefont {Erik P.
  A.~M.}\ \bibnamefont {Bakkers}}, \bibinfo {author} {\bibfnamefont {Michael}\
  \bibnamefont {Wimmer}}, \ and\ \bibinfo {author} {\bibfnamefont {Leo~P.}\
  \bibnamefont {Kouwenhoven}},\ }\bibfield  {title} {\enquote {\bibinfo {title}
  {Realization of a minimal kitaev chain in coupled quantum dots},}\ }\href
  {\doibase 10.1038/s41586-022-05585-1} {\bibfield  {journal} {\bibinfo
  {journal} {Nature}\ }\textbf {\bibinfo {volume} {614}},\ \bibinfo {pages}
  {445--450} (\bibinfo {year} {2023})}\BibitemShut {NoStop}%
\bibitem [{\citenamefont {Sela}\ \emph {et~al.}(2011)\citenamefont {Sela},
  \citenamefont {Altland},\ and\ \citenamefont {Rosch}}]{Sela2011}%
  \BibitemOpen
  \bibfield  {author} {\bibinfo {author} {\bibfnamefont {Eran}\ \bibnamefont
  {Sela}}, \bibinfo {author} {\bibfnamefont {Alexander}\ \bibnamefont
  {Altland}}, \ and\ \bibinfo {author} {\bibfnamefont {Achim}\ \bibnamefont
  {Rosch}},\ }\bibfield  {title} {\enquote {\bibinfo {title} {Majorana fermions
  in strongly interacting helical liquids},}\ }\href {\doibase
  10.1103/PhysRevB.84.085114} {\bibfield  {journal} {\bibinfo  {journal} {Phys.
  Rev. B}\ }\textbf {\bibinfo {volume} {84}},\ \bibinfo {pages} {085114}
  (\bibinfo {year} {2011})}\BibitemShut {NoStop}%
\bibitem [{\citenamefont {Liu}\ \emph {et~al.}(2021)\citenamefont {Liu},
  \citenamefont {Xu},\ and\ \citenamefont {Li}}]{Liu2021}%
  \BibitemOpen
  \bibfield  {author} {\bibinfo {author} {\bibfnamefont {Yu-Guo}\ \bibnamefont
  {Liu}}, \bibinfo {author} {\bibfnamefont {Lu}~\bibnamefont {Xu}}, \ and\
  \bibinfo {author} {\bibfnamefont {Zhi}\ \bibnamefont {Li}},\ }\bibfield
  {title} {\enquote {\bibinfo {title} {Quantum phase transition in a
  non-hermitian {XY} spin chain with global complex transverse field},}\ }\href
  {\doibase 10.1088/1361-648x/ac00dd} {\bibfield  {journal} {\bibinfo
  {journal} {Journal of Physics: Condensed Matter}\ }\textbf {\bibinfo {volume}
  {33}},\ \bibinfo {pages} {295401} (\bibinfo {year} {2021})}\BibitemShut
  {NoStop}%
\bibitem [{\citenamefont {Bi}\ \emph {et~al.}(2021)\citenamefont {Bi},
  \citenamefont {He},\ and\ \citenamefont {Li}}]{Bi2021}%
  \BibitemOpen
  \bibfield  {author} {\bibinfo {author} {\bibfnamefont {Shihao}\ \bibnamefont
  {Bi}}, \bibinfo {author} {\bibfnamefont {Yan}\ \bibnamefont {He}}, \ and\
  \bibinfo {author} {\bibfnamefont {Peng}\ \bibnamefont {Li}},\ }\bibfield
  {title} {\enquote {\bibinfo {title} {Ring-frustrated non-hermitian {XY}
  model},}\ }\href {\doibase 10.1016/j.physleta.2021.127208} {\bibfield
  {journal} {\bibinfo  {journal} {Physics Letters A}\ }\textbf {\bibinfo
  {volume} {395}},\ \bibinfo {pages} {127208} (\bibinfo {year}
  {2021})}\BibitemShut {NoStop}%
\bibitem [{\citenamefont {Schultz}\ \emph {et~al.}(1964)\citenamefont
  {Schultz}, \citenamefont {Mattis},\ and\ \citenamefont {Lieb}}]{Schultz1964}%
  \BibitemOpen
  \bibfield  {author} {\bibinfo {author} {\bibfnamefont {T.~D.}\ \bibnamefont
  {Schultz}}, \bibinfo {author} {\bibfnamefont {D.~C.}\ \bibnamefont {Mattis}},
  \ and\ \bibinfo {author} {\bibfnamefont {E.~H.}\ \bibnamefont {Lieb}},\
  }\bibfield  {title} {\enquote {\bibinfo {title} {Two-dimensional ising model
  as a soluble problem of many fermions},}\ }\href {\doibase
  10.1103/RevModPhys.36.856} {\bibfield  {journal} {\bibinfo  {journal} {Rev.
  Mod. Phys.}\ }\textbf {\bibinfo {volume} {36}},\ \bibinfo {pages} {856--871}
  (\bibinfo {year} {1964})}\BibitemShut {NoStop}%
\bibitem [{\citenamefont {Ding}\ and\ \citenamefont {Zhong}(2021)}]{Ding2021}%
  \BibitemOpen
  \bibfield  {author} {\bibinfo {author} {\bibfnamefont {Lin-Jie}\ \bibnamefont
  {Ding}}\ and\ \bibinfo {author} {\bibfnamefont {Yuan}\ \bibnamefont
  {Zhong}},\ }\bibfield  {title} {\enquote {\bibinfo {title} {Gruneisen ratio
  quest for self-duality of quantum criticality in a spin-1/2 {XY} chain with
  dzyaloshinskii{\textendash}moriya interaction},}\ }\href {\doibase
  10.1088/1572-9494/ac061d} {\bibfield  {journal} {\bibinfo  {journal}
  {Communications in Theoretical Physics}\ }\textbf {\bibinfo {volume} {73}},\
  \bibinfo {pages} {095701} (\bibinfo {year} {2021})}\BibitemShut {NoStop}%
\bibitem [{\citenamefont {Wada}\ \emph {et~al.}(2021)\citenamefont {Wada},
  \citenamefont {Sugimoto},\ and\ \citenamefont {Tohyama}}]{Wada2021}%
  \BibitemOpen
  \bibfield  {author} {\bibinfo {author} {\bibfnamefont {Kazuhiro}\
  \bibnamefont {Wada}}, \bibinfo {author} {\bibfnamefont {Takanori}\
  \bibnamefont {Sugimoto}}, \ and\ \bibinfo {author} {\bibfnamefont {Takami}\
  \bibnamefont {Tohyama}},\ }\bibfield  {title} {\enquote {\bibinfo {title}
  {{Coexistence of strong and weak Majorana zero modes in an anisotropic XY
  spin chain with second-neighbor interactions}},}\ }\href {\doibase
  10.1103/PhysRevB.104.075119} {\bibfield  {journal} {\bibinfo  {journal}
  {Physical Review B}\ }\textbf {\bibinfo {volume} {104}},\ \bibinfo {pages}
  {1--12} (\bibinfo {year} {2021})},\ \Eprint {http://arxiv.org/abs/2103.12281}
  {arXiv:2103.12281} \BibitemShut {NoStop}%
\bibitem [{dmr()}]{dmrgpy}%
  \BibitemOpen
  \href@noop {} {\bibinfo  {journal} {\mbox{DMRGpy Library}
  https://github.com/joselado/dmrgpy}\ }\BibitemShut {NoStop}%
\bibitem [{ITe()}]{ITensor}%
  \BibitemOpen
\bibfield  {journal} {  }\href@noop {} {\bibinfo  {journal} {\mbox{ITensor
  Library} http://itensor.org}\ }\BibitemShut {NoStop}%
\bibitem [{\citenamefont {Fishman}\ \emph {et~al.}(2022)\citenamefont
  {Fishman}, \citenamefont {White},\ and\ \citenamefont
  {Stoudenmire}}]{Fishman2022}%
  \BibitemOpen
\bibfield  {journal} {  }\bibfield  {author} {\bibinfo {author} {\bibfnamefont
  {Matthew}\ \bibnamefont {Fishman}}, \bibinfo {author} {\bibfnamefont
  {Steven~R.}\ \bibnamefont {White}}, \ and\ \bibinfo {author} {\bibfnamefont
  {E.~Miles}\ \bibnamefont {Stoudenmire}},\ }\bibfield  {title} {\enquote
  {\bibinfo {title} {{The ITensor Software Library for Tensor Network
  Calculations}},}\ }\href {\doibase 10.21468/SciPostPhysCodeb.4} {\bibfield
  {journal} {\bibinfo  {journal} {SciPost Phys. Codebases}\ ,\ \bibinfo {pages}
  {4}} (\bibinfo {year} {2022})}\BibitemShut {NoStop}%
\end{thebibliography}%

\end{document}